\documentclass[a4paper]{article}

\usepackage{stdformat}
\usepackage{subfigure}
\usepackage[font=small, labelfont=bf]{caption}
\usepackage{authblk}

\usepackage[linesnumbered,ruled]{algorithm2e}
\SetAlCapNameFnt{\small}
\SetAlCapFnt{\small}

\usepackage{url}

\usepackage[all]{nowidow} % tex.stackexchange.com/a/30409

\title{A simulation framework of procurement operations in the container logistics industry}

\author[1,3]{George Vassos}
\author[1]{Klaus K. Holst}
\author[2,3]{Pierre Pinson}
\author[3]{Richard~M. Lusby}
\affil[1]{Research and Development, A.P. Moller - Maersk}
\affil[2]{Dyson School of Design Engineering, Imperial College London}
\affil[3]{Department of Technology, Management and Economics, Technical University of Denmark}

\begin{document}
\maketitle

\vspace{5mm}

\begin{abstract}
    \noindent This study proposes a simulation framework of procurement operations in the container logistics industry that can support the development of dynamic procurement strategies. The idea is inspired by the success of Passenger Origin-Destination Simulator (PODS) in the field of airline revenue management. By and large, research in procurement has focused on the optimisation of purchasing decisions, i.e., when-to-order and supplier selection, but a principled approach to procurement operations is lacking. We fill this gap by developing a probabilistic model of a procurement system. A discrete-event simulation logic is used to drive the evolution of the system. In a small case study, we use the simulation to deliver insights by comparing different supplier selection policies in a dynamic spot market environment. Policies based on contextual multi-armed bandits are seen to be robust to limited access to the information that determines the distribution of the outcome. This paper provides a pool of modelling ideas for simulation and observational studies. Moreover, the probabilistic formulation paves the way for advanced machine learning techniques and data-driven optimisation in procurement.
\end{abstract}

\section{Introduction}\label{sec:1}

\begin{figure}
    \centering
    \includegraphics[width=\textwidth]{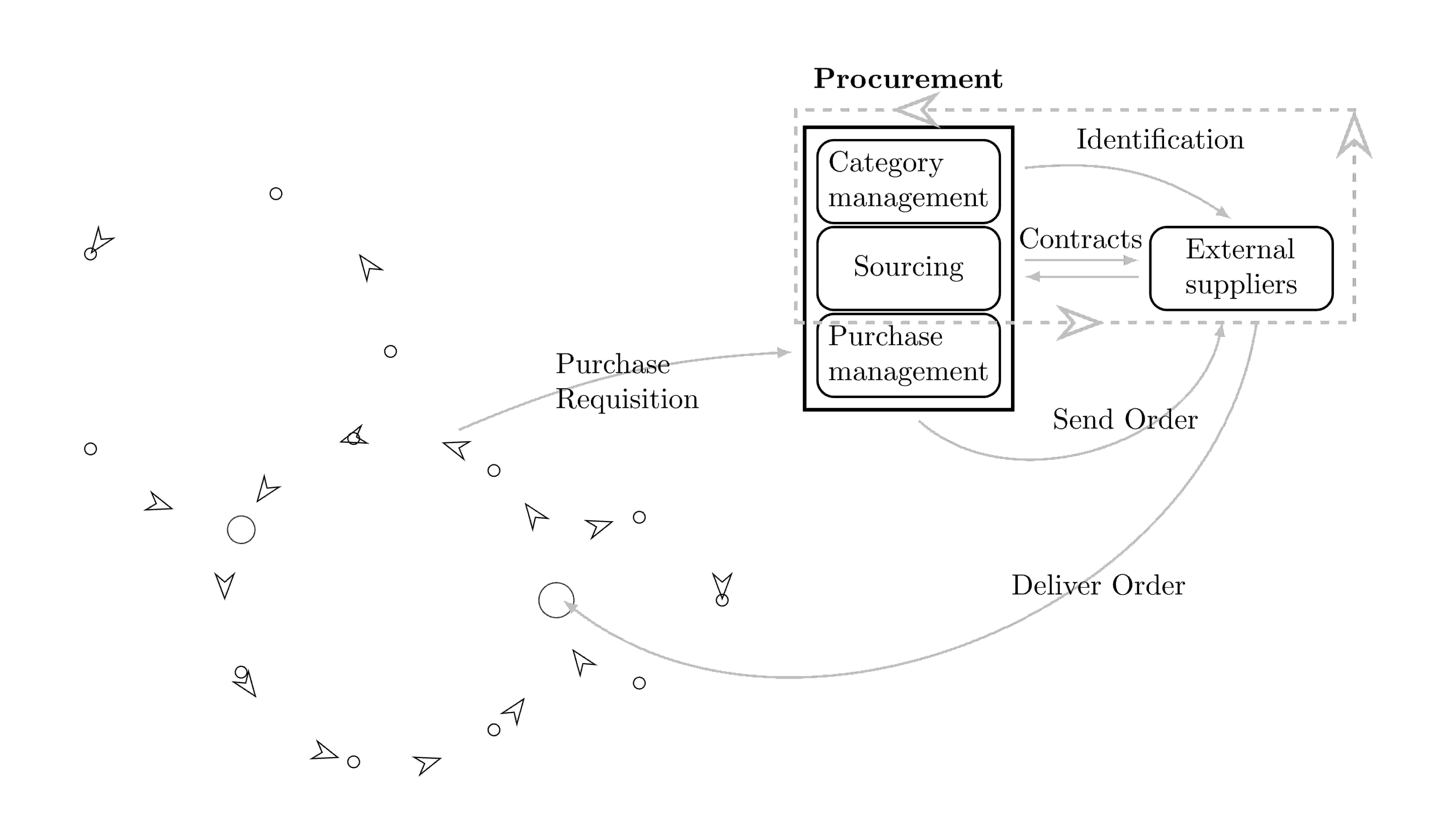}
    \caption{Procurement in the container logistics industry: circles and arrow heads, on the left side, represent operational sites and transportation units or vice versa. The sites generate demand, i.e., product and service requests, that is sent to the operational layer of the procurement organisation where purchase orders are executed. The higher layers, that is, category management and sourcing are responsible for developing and implementing the business policies to guide the decisions of purchase management.}
    \label{fig:procorgex}
\end{figure}

The procurement business is important to companies with widespread or global supply chains, where, naturally, demand for products and services is generated daily from business operations. Therefore, a procurement organisation has to be in place to ensure such demand is met in a structured fashion. The procurement organisation is a three-layer hierarchical entity, within a company, with the purpose of carrying out the procurement enterprise. On the top layer, category management identifies categories of products and services and develops strategies for establishing framework agreements (contracts) with suppliers. On the middle layer, sourcing must implement the strategies to consolidate a supplier portfolio with contract and spot market options. On the bottom layer, purchase managers execute the strategy by allocating demand to suppliers in daily operations. This generic structure of procurement is depicted on the right side of Figure~\ref{fig:procorgex}. In the container logistics industry, there is customer demand for container transportation. A container logistics company utilises a fleet of transportation units that can be owned by the company or by service providers in a contract-spot market. Procurement maintains a contract-spot portfolio of container transportation suppliers as well as suppliers that can satisfy demand for the operation of company-owned assets. Figure~\ref{fig:procorgex} illustrates the structure of the procurement business in container logistics. Knowledge and insights about the nature of operations are core determinants of a procurement strategy.

\begin{figure}
    \centering
    \includegraphics[width=0.55\textwidth]{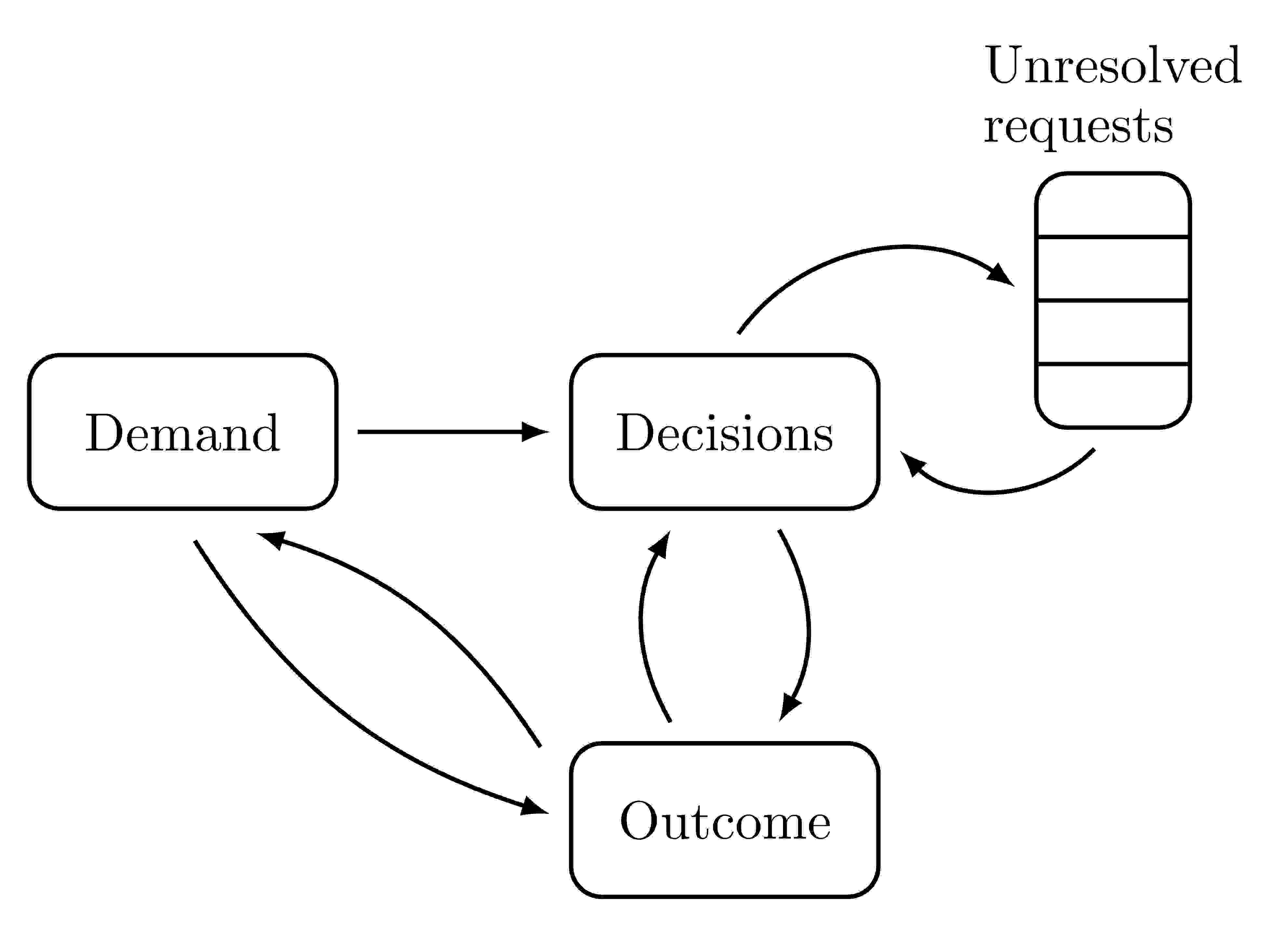}
    \caption{Generic architecture of the simulation logic including the three core components, namely, the demand, decision, and outcome mechanisms, the auxiliary component where the pending requests are maintained, and the interactions between them represented by the arrows. Notice how the demand may only be impacted by the decisions through the outcome.}
    \label{fig:logic}
\end{figure}

% Add something here on the use case of PODS and our source code.
In a typical procurement process, strategic and tactical planning are made in a static fashion, similar to airlines before the development and success of the Passenger Origin-Destination Simulator (PODS), in the sense that they rely on a long-term forecast of the demand, which is then fed into optimisation models to develop operational policies over annual or quarterly horizons \citep[~Chapter 3]{fry2015a}. PODS is a tool for revenue management developed by Boeing with a simulation model of customer demand and few options for the number of airlines and network competition. Similar to PODS, the proposed framework can enable a transition towards more flexible solutions where the effect of strategic and tactical decisions is continuously informed and policies are dynamically updated in light of new data. This study develops a probabilistic formulation of the mechanism underlying the evolution of a procurement system where in the ambient scheme demand, decisions, and outcome (welfare measures) are interdependent, as shown in Figure~\ref{fig:logic}. We believe this study can spark future research in procurement, as the idea behind PODS has led to a series of concomitant studies \citep{fry2015a,wittman2016a,weatherford2017a,wang2021a,szyma2021a}, leading to a valuable product for revenue optimisation.

Our goal is to advance a joint probabilistic model of operations and demand in procurement, fostering the use of advanced machine learning techniques in optimisation which is currently trending in the Operations Research (OR) field \citep{lodi2020}. \cite{Chilmon2020} and \cite{Ha2018} argue on the importance of a unified framework for procurement operations in which demand and operations are interconnected. The simultaneous optimisation of all model parameters is very likely to be intractable even for simple simulation configurations; however, extensive control over the evolution of the system can help formulate more concise optimisation problems than those under the usual untestable exogeneity assumptions on demand \citep{Zheng2021,Zhang2013} or spot market \citep{Nie2017,Hu2012}. The state of the procurement system is determined by historical (demand, decision, outcome) records, exogenous effects, and the list of unresolved requests. On a given day, purchase management will decide whether or not and to whom to send purchase orders given the present state of the system, which is in turn affected by the decisions taken.

A case study is carried out to illustrate how the simulation can be used to compare different supplier selection policies in a dynamic spot market environment. Supplier selection is the primary operational task in procurement where the business strategy is executed to realise the expectations of category management. The utility of the spot market is twofold: first, it is a safety buffer against disruptions in demand and helps to avoid supplier lock-in \citep{Hamdi2018}, and, second, it enables the exploration of new collaboration opportunities. Contextual multi-armed bandits are known to have robust performance in non-stationary environments \citep{Sutton1998}. Our insights indicate that in a spot market situation contextual multi-armed bandits outperform static supplier selection policies with the same knowledge of the outcome distribution. Hence, such models can be employed to perform efficient supplier selection or exploration of new opportunities in the spot market and continuously inform a more dynamic business strategy.

This paper is structured as follows. Section~\ref{sec:lr} is a review of state-of-the-art literature on modern approaches to procurement and the need for a new paradigm. Section~\ref{sec:outcome}, provides a unified probabilistic model and the set of assumptions under which it describes the data generating mechanism in the procurement business. An outline of a discrete-event simulation \citep{law2000des} design is included to inform software implementations. Section~\ref{sec:propmod}, proposes specific expressions and algorithms for the components of the probabilistic model, and we showcase the usage of our proposed framework in an example in Section \ref{sec:cstud}. Finally, Section~\ref{sec:concl} discusses future research perspectives.

\section{Literature review}\label{sec:lr}

% Importance of the procurement business and insights in procurement 

The significance of the procurement organisation to modern multinational corporations has been re-evaluated in recent years to accommodate the complexity of supply chain planning in the global environment \citep{Hamdi2018}. Strategic collaborations with suppliers and the establishment of strong supplier relationships are critical to the efficiency of daily operations. The work by \cite{Martinez2005a} shows the link between the choice of a supply contract portfolio and the optimisation of procurement operations. \cite{Hamdi2018} explain the important role of supplier selection in supply chain risk mitigation.

% Current focus of literature and shift to new paradigm

By and large, research in the field of procurement and supply chain planning has focused on stochastic and robust optimisation but the field of OR \citep{govindan2018}. These traditional OR approaches often rely on long-term forecasts of the demand and assumptions about the market environment that are postulates and do not emerge from data modelling. Limitations of such approaches have been outlined by \cite{Abbasi2020}, \cite{powell2019}, and \cite{Shahriari2016}. \cite{dai2020} criticised the use of long-term forecasts in the development of procurement strategies. The situation is similar to that with the airline revenue management business that led to the development of PODS which has been a key lever in the transition from static to dynamic revenue management in the airline industry \citep[~Chapter 3]{fry2015a}.

The recent advances in machine learning and data-driven optimisation promote a new paradigm in procurement research to enable the transition to dynamic strategies. A trend towards introducing machine learning in supply chain management research within OR has been highlighted by \cite{lodi2020}, \cite{Larsen2018}, and \cite{Nguyen2018}. Policy learning under uncertainty in a Markov decision process (MDP) framework has been attempted on similar applications \citep{powell2019,haehl2018a}. Moreover, concerns on the causal interpretation of OR models have been raised \citep{bertsimas2020a,bertsimas2017a}. In a systematic literature review, \cite{Chilmon2020} pointed out the great utility of simulation methods in modelling complex systems such as an end-to-end supply chain (E2ESC). They highlighted the absence of a principled framework, i.e., of generic rules, concepts, and processes, which can provide regularity for the future efforts in modelling complex systems, and emphasised the importance of having a theoretical model for real world problems, a point also made by \cite{Ha2018}, as well as the scarcity of such efforts \citep{Chilmon2020}.

% Importance of supplier selection that is the topic of the case study

Supplier or vendor selection is one of the operational problems that falls within the scope of supply chain planning. Supplier selection can be understood as the execution of a procurement strategy in daily operations. \cite{govindan2018} stress the potential of multidisciplinary research to improve supply chain planning in uncertain environments. \cite{Nguyen2018} found a growing interest in data-driven decision-making in supply chain management (SCM) in recent years. They suggested that data-driven approaches have attracted great interest among researchers in confronting the challenges of SCM. % Suggest the use of contextual multi-armed bandits as a data-driven approach.

%%%%%%%%%%%%%%%%%%%%%%%%%%%%%%%%%%%%%%%%

% Exhaustive analysis of related Production Research publications

Important problems of procurement such as the identification of optimal supplier selection policies have been studied in isolation taking approaches such as fuzzy multi-criteria decision-making (MCMD), game theory, swarm-based heuristics, robust optimisation, or fuzzy goal programming \citep{Khemiri2017,Chen2015,Yoon2020,Prince2013,Pal2011,Kanyalkar2010,Torabi2009}. Such studies focused on obtaining insights on very specific study designs and do not provide a portable logic that can be readily utilised by new researchers. Interest has also been shown in procurement policies under uncertainty, usually on price or demand, \citep{dai2020,Bollapragada2015,Xie2013,Hegedus2001}; however, the standard practice is to postulate a probability distribution on the price or demand without any insights on the underlying generating mechanism. Data-driven approaches to procurement are few and fairly recent \citep{Zhang2022,Mogale2020,akcay2017} and have also been anticipated due to the rise of e-procurement systems \citep{Chibani2018,Oh2014,Chan2011,Talluri2007}. Case-specific simulation models have been developed for the investigation of particular research inquiries \citep{Tao2020,He2016,Johnson2013} but they are not generalisable to confront a wider variety of procurement problems. Interestingly, simulation alternatives have been employed to enable the comparison of procurement policies \citep{Mula2013}. Procurement has also been considered jointly with other processes in a broader supply chain planning context \citep{Karabag2022,Yaghin2021,Reiner2014}. Joint optimisation at such scale could easily become intractable. The complexity burden could be mitigated with the use of a simulation model of procurement. Buyer-supplier interaction is a core topic in procurement \citep{Ghadimi2016} and also a focal point of the proposed simulation model. Other cases that are of interest in recent literature on procurement are concerned with green or sustainable procurement \citep{Niu2021,Ghadge2019,Dey2019} and can also benefit from using a simulation.

% Explanation of the proposed model methodologies

This study proposes a simulation framework of procurement operations that is founded on a probabilistic model that captures the dynamics between operations and demand. On the highest level, the model consists of demand, decision, and outcome components with an interdependence structure. Furthermore, a set of configurable specifications allows great flexibility when simulating a variety of scenarios, and enables both simulation-based optimisation and data-driven approaches. Demand is considered in the form of purchase requisitions which are common in the procurement business and flexible to configure. The important operational decisions in procurement are: (1) when to place an order, (2) supplier selection, and (3) demand allocation \citep{Sun2022}. Attempts to model (1) as an optimal stopping problem are made by \cite{xie2013a} and \cite{fontes2008}. Research in (2) throughout a relatively long period of time has pointed out cost, lead or delivery time, and quality as the main drivers of the business \citep{rao2017a,weber19912,dickson1966}. The problem of demand allocation in procurement is linked to compliance with existing supply contracts. The proposed simulation framework can be configured to generate insights in this situation.

The novelty of this study is a formulation of the procurement system that avoids any independence assumptions between operations and demand or spot prices and makes no assumptions on the functional form of the random variables that describe the procurement system. Previous studies have only treated the case of independent demand, spot market, and procurement operations under regularity assumptions that allowed for convenient solutions at the expense of generality.

\section{Mathematical framework and simulation logic}\label{sec:outcome}

In this section, we propose a joint probabilistic model of procurement operations, decision-making and outcome mechanisms, and demand. We introduce a model of the evolution of procurement operations which enables a systematic representation of the problem.

Let $B_{\ell}$ denote the demand information available at calendar day $\ell$, with $\ell=1,\dots,\tau$ enumerating $\tau$ calendar days, and $m=1,\dots,M_{\ell}$, where $M_{\ell}$ is the total number of entries in all requisitions in $B_{\ell}$. We use the term line item to refer to a single entry, i.e., request for a product, in a requisition. Let $\mathcal{D}_{\ell}$ denote the set of suppliers who are relevant at time $\ell$, and $D_{\ell m}\in\{(0,\emptyset),(1,1),\dots,(1,|\mathcal{D}_{\ell}|)\}$ the relative decision variable. That is, at any calendar day $\ell$, the purchase management needs to decide whether to place a purchase order, for line item $m$, or not. If the decision is to order it, then one out of $|\mathcal{D}_{\ell}|$ distinct supplier firms must be selected to send the order to. We let $Y_{\ell m}$ denote the response variable, i.e., cost, delivery time, and quality, relative to line item $m$ at calendar day $\ell$.
\begin{example}
    Suppose we observe for $2$ days a procurement system with three products $\{1,2,3\}$ that belong to a single category and two suppliers $\{1,2\}$ that can provide all three products. Following the logic of the proposed data generating mechanism, we illustrate below a potential observation from this system. Notice that $b_{1}$ is a purchase requisition that contains two line items $b_{11}$ and $b_{13}$ and similar for $b_{2}$.
    \begin{equation*}
    \begin{matrix}
       \text{Product id} & \vrule & b_{1} & d_{1} & y_{1} & b_{2} & d_{2} & y_{2} \\
       \midrule
       1 & \vrule & b_{11} & (0,\emptyset) & \emptyset & b_{21} & (1,1) & y_{21} \\
       2 & \vrule & \emptyset & \emptyset & \emptyset & b_{22} & (1,1) & y_{22} \\
       3 & \vrule & b_{13} & (1,2) & y_{13} & \emptyset & \emptyset & \emptyset \\
    \end{matrix}
    \end{equation*}
\end{example}
A realisation of the system over a period of $\tau$ calendar days will have the structure
\begin{equation}\label{eq:longobs}
    (b_{1},d_{1},y_{1}),\dots,(b_{\tau},d_{\tau},y_{\tau})
\end{equation}
which is a draw from the probability distribution, also called mechanism, with density function that can be factorised as
\begin{equation}\label{eq:jointmechlong}
    \prod_{\ell=1}^{\tau}p(b_{\ell}\,|\,\overline{y}_{\ell-1},\overline{d}_{\ell-1},\overline{b}_{\ell-1})\,p(d_{\ell}\,|\,\overline{y}_{\ell-1},\overline{d}_{\ell-1},\overline{b}_{\ell})\,p(y_{\ell}\,|\,\overline{y}_{\ell-1},\overline{d}_{\ell},\overline{b}_{\ell})
\end{equation}
where $\overline{b}_{\ell}\doteq(b_{1},\dots,b_{\ell})$, $\overline{d}_{\ell}\doteq(d_{1},\dots,d_{\ell})$ and $\overline{y}_{\ell}\doteq(y_{1},\dots,y_{\ell})$, and we use the convention that $\overline{b}_{0}=\overline{d}_{0}=\overline{y}_{0}=\emptyset$. Letting $b_{\ell}^{-}\doteq(\overline{y}_{\ell-1},\overline{d}_{\ell-1},\overline{b}_{\ell-1})$, $d_{\ell}^{-}\doteq(\overline{y}_{\ell-1},\overline{d}_{\ell-1},\overline{b}_{\ell})$, and $y_{\ell}^{-}\doteq(\overline{y}_{\ell-1},\overline{d}_{\ell},\overline{b}_{\ell})$ we can express the probability density function more compactly as
\begin{equation}\label{eq:jointmech}
    \prod_{\ell=1}^{\tau}p(b_{\ell}\,|\,b_{\ell}^{-})\,p(d_{\ell}\,|\,d_{\ell}^{-})\,p(y_{\ell}\,|\,y_{\ell}^{-})
\end{equation}

\begin{figure}
    \centering
    \includegraphics{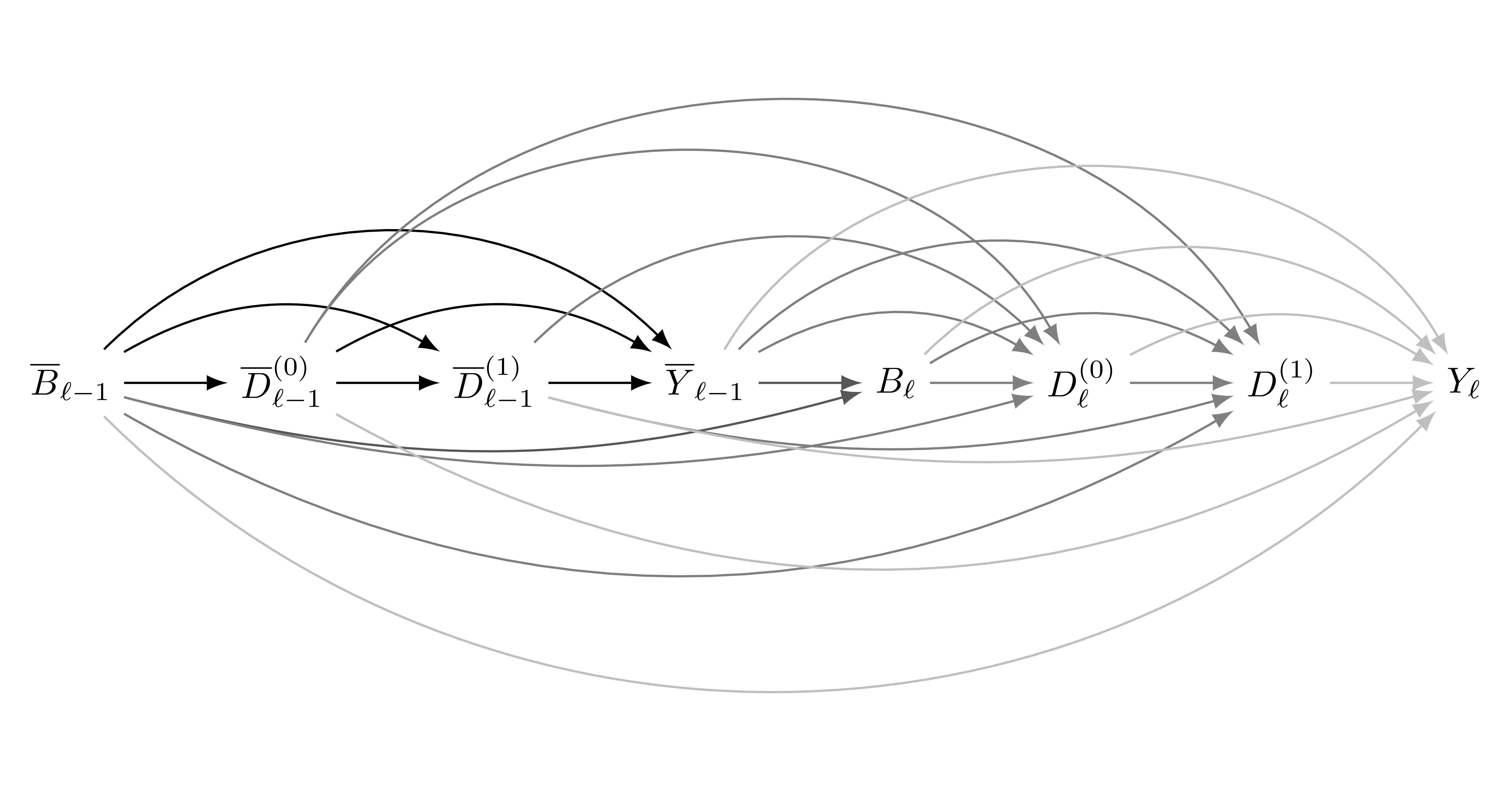}
    \caption{A directed acyclic graph (DAG) that illustrates the dynamics in the model under the assumption that there is no unobserved confounding so that any arrow represents a direct cause-effect relationship. A DAG is useful by way of communicating the set of cause-effect relationships that are at work in a particular simulation scenario.}
    \label{fig:desgop-dag}
\end{figure}

A mechanism encapsulates the entirety of cause and effect relationships that govern the evolution of a system. Directed acyclic graphs (DAGs) are common devices, in the theory coarsely termed Causality or Causal Inference \citep{hernan2020a,peters2017}, used to represent structural causal models; in the context of this study, a structural causal model can be seen as a collection of probability distributions having a density of the sort of \eqref{eq:jointmech}. An arrow in a DAG indicates a cause-effect relationship between the connected nodes and is always directed from left to right. Figure~\ref{fig:desgop-dag} attempts an economic illustration of the longitudinal observation structure \eqref{eq:longobs} aiming to highlight the working cause-effect relationships at time $\ell$. Figure~\ref{fig:desgop-dag} identifies the model which includes a conditional independence assumption.

\begin{assumption}[Conditional independence]\label{asm:ci}
    Demand at time $\ell$ is independent of past decisions given the historical demand and outcome information, i.e.,
    \begin{equation}
        B_{\ell}\idp\overline{D}_{\ell-1}\,|\,\overline{B}_{\ell-1},\overline{Y}_{\ell-1}
    \end{equation}
\end{assumption}

The assumption asserts that past decisions can affect demand only through the outcome, and is the initial design assumption presented in Figure~\ref{fig:logic}. The reader can verify the absence of arrows from $\overline{D}_{\ell-1}^{(0)}$ and $\overline{D}_{\ell-1}^{(1)}$ to $B_{\ell}$ in Figure~\ref{fig:desgop-dag}, and notice the flexibility of the proposed model in terms of capturing cause-effect relationships. Under the simplifying Assumption \ref{asm:ci}, Equation~\eqref{eq:jointmech} remains a valid expression of the probability density function with the difference that $b_{\ell}^{-}\doteq(\overline{y}_{\ell-1},\overline{b}_{\ell-1})$.

\begin{figure}
  \centering
  \subfigure[]{%
  \resizebox*{7cm}{!}{\includegraphics{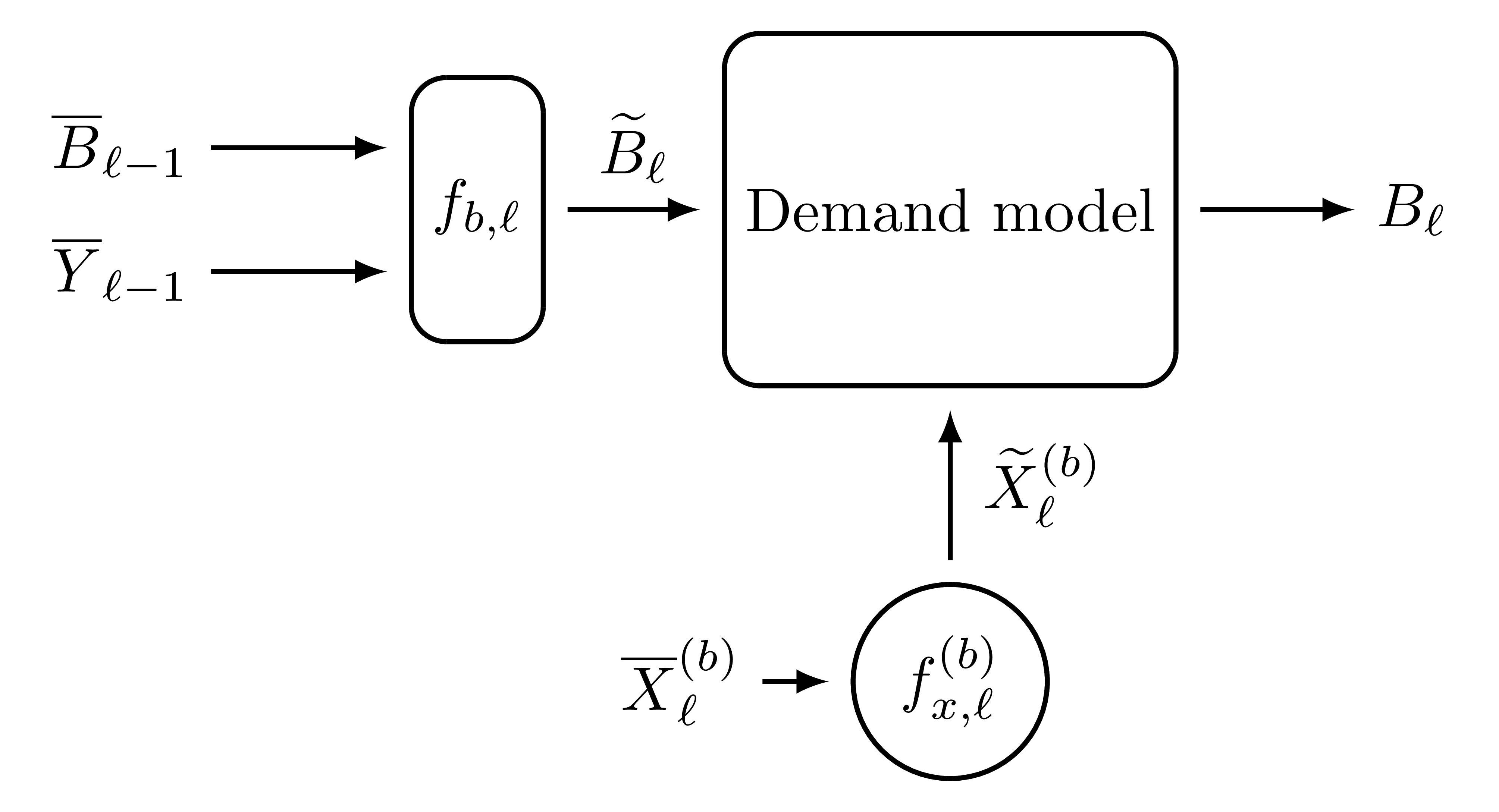}}}\hspace{5pt}
  \subfigure[]{%
  \resizebox*{7cm}{!}{\includegraphics{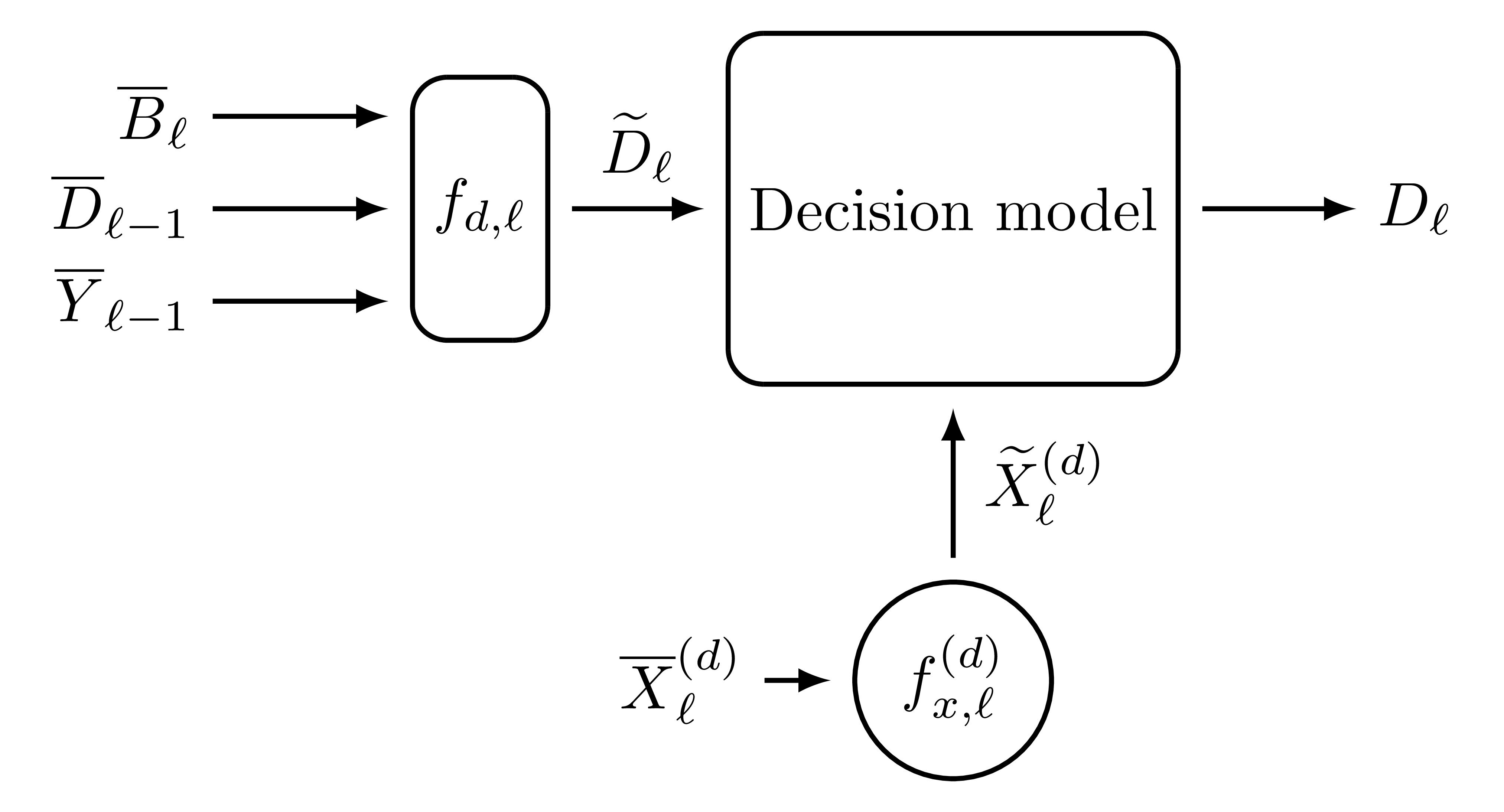}}}
  \subfigure[]{%
  \resizebox*{7cm}{!}{\includegraphics{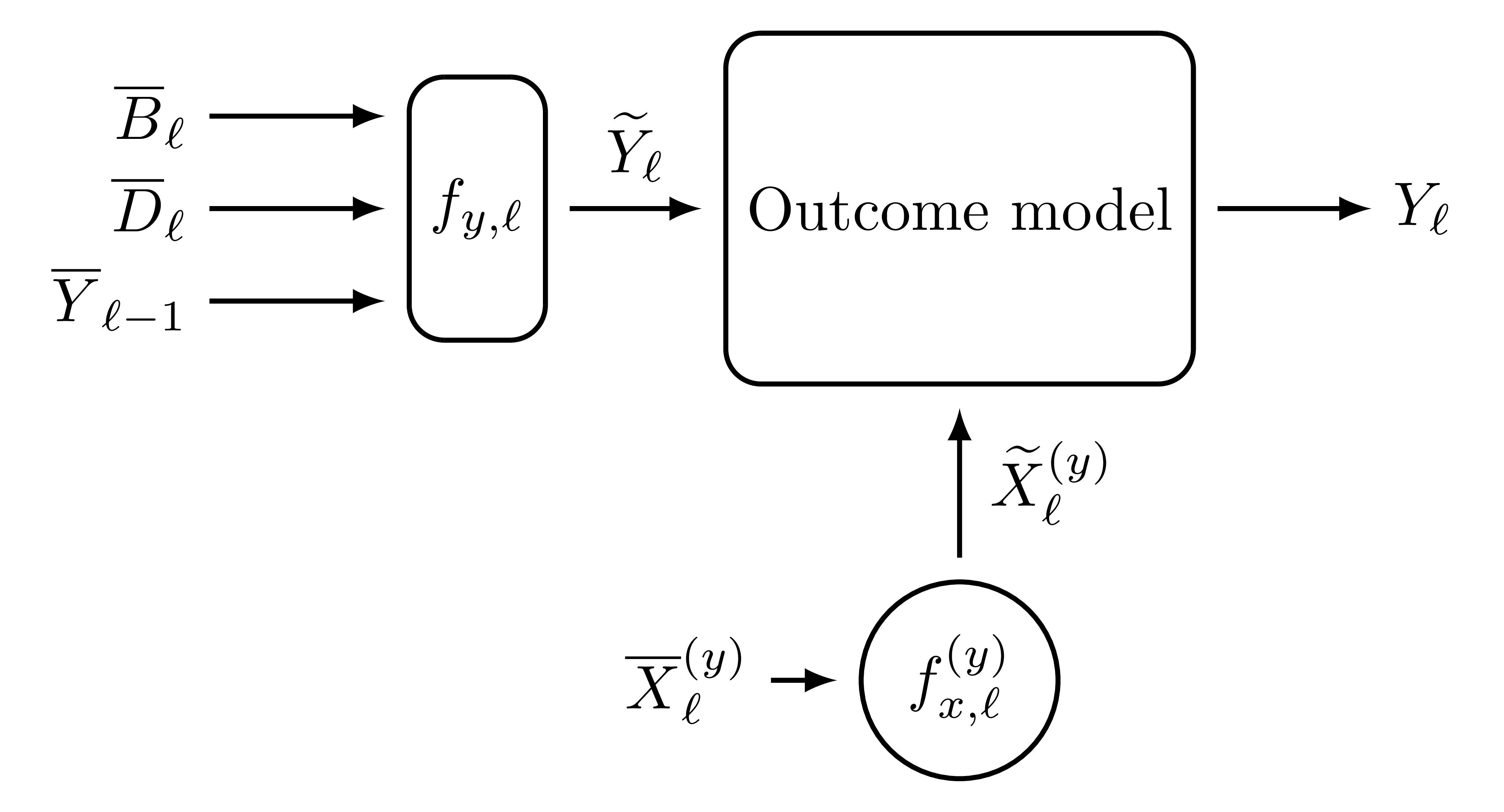}}}
  \caption{Graphical input/output illustration of the (a) demand model, (b) decision model, and the (c) outcome model. Each model takes as input a number of features generated from the endogenous variables of the relative mechanism as reported in \eqref{eq:jointmech} as well as additional features constructed from exogenous covariates and generates the corresponding output variable.}
  \label{fig:models}
\end{figure}

In the general case, we would like to generate a number of features from historical data, that is, to construct a design using transformations $B_{\ell}^{-}\mapsto f_{b,\ell}(B_{\ell}^{-})\doteq\widetilde{B}_{\ell}$, $D_{\ell}^{-}\mapsto f_{d,\ell}(D_{\ell}^{-})\doteq \widetilde{D}_{\ell}$, and $Y_{\ell}^{-}\mapsto f_{y,\ell}(Y_{\ell}^{-})\doteq \widetilde{Y}_{\ell}$, and, potentially, introduce additional exogenous covariates
\begin{equation*}
    \left(\overline{X}_{\ell}^{(b)},\overline{X}_{\ell}^{(d)},\overline{X}_{\ell}^{(y)}\right)\mapsto\left(f_{x,\ell}^{(b)}(\overline{X}_{\ell}^{(b)}),f_{x,\ell}^{(d)}(\overline{X}_{\ell}^{(d)}),f_{x,\ell}^{(y)}(\overline{X}_{\ell}^{(y)})\right)\doteq\left(\widetilde{X}_{\ell}^{(b)},\widetilde{X}_{\ell}^{(d)},\widetilde{X}_{\ell}^{(y)}\right)
\end{equation*}
so that $(\widetilde{B}_{\ell},\widetilde{X}_{\ell}^{(b)})$ can serve as input to the demand model for generating the next demand instance $B_{\ell}$, $(\widetilde{D}_{\ell},\widetilde{X}_{\ell}^{(d)})$ can be used in the decision model to generate $D_{\ell}$, and $(\widetilde{Y}_{\ell},\widetilde{X}_{\ell}^{(y)})$ would be the input to the outcome model for obtaining $Y_{\ell}$. Figure~\ref{fig:models} demonstrates the logic in three simple diagrams. In the following sections, we describe in more detail the demand, decision, and outcome models that we have introduced.

\subsection{Demand model}

We propose a two-stage model for the generation of purchase requisitions. First, a recurrent event process, modelled with an intensity function, triggers the generation of a requisition; we may refer to the exact time when a generation is triggered as an event time. Furthermore, we argue that the nature of a procurement organisation allows for an independence assumption on the demand between categories so that we can focus on one category without loss of generality. Second, once a requisition generation has been triggered, we describe the selection of line items with a sequential decision model.

\subsubsection{Time-to-request model}

A procurement setting involves operational sites that generate recurrent purchase requisitions for different categories of products. One of the strategic tasks of a procurement organisation, as mentioned in Section~\ref{sec:1}, is to segregate the products into categories so that each category can be treated independently. Thus, we can naturally make an independence assumption which allows us to consider a single category as our point of reference and readily generalise our analysis to all categories.
\begin{assumption}[Independent categories]
    Demand is independent across categories of products.
\end{assumption}
Suppose at time $t$, $N_{t}$ requisitions are sent to procurement from all operations so that, in continuous time, there is a counting process $(N_{t}:t\ge0)$ at work. Under the temporal ordering, it is possible to introduce the demand process $(B_{t}:t\ge0)$ as well as additional exogenous processes associated with the counting process. The sequence of event times, at which requisitions are generated, is guided by a recurrent event process which, in line with the literature on event history analysis \citep{cook2007a,aalen2008a}, is modelled using an intensity function $\lambda(\,\cdot\,|\,H_{t})$, where $H_{t}$ contains historical counts $(N_{s}:0\le s<t)$ and information exogenous to the counting process, namely, the covariate process $(X_{t}^{(b)}:t\ge0)$ and historical process $B_{t}^{-}\doteq((B_{s},Y_{s}):0\le s<t)$.

Event times take values in the event time scale induced by $\lambda(\,\cdot\,|\,H_{t})$ which is, in our case, indexed by $t$ and it is the calendar scale but can be finer than days, e.g., hours. \cite{cook2007a} define the intensity function, that is, the probability that a requisition is generated exactly at time $t$, as
\begin{equation}
    \lambda(t\,|\,H_{t})\doteq\lim_{\Delta t\to0}\frac{P(\Delta N_{t}=1\,|\,H_{t})}{\Delta t}
\end{equation}
and can be used to model the generation of requisitions for some category of products at an operational site. When modelling a recurrent event process with an intensity function it is customary to assume that no more than one requisition can be generated at a particular time instance \citep{cook2007a,aalen2008a}. In this study, it is easy to force the veracity of this assumption in the rare situation that two events are generated simultaneously by assigning them a random order and adding a very small increment to event(s) in later order.

In this study, an event time can always be identified with an operational site $i=1,\dots,n$ and the number of requisitions generated at that site $n_{i}$; in the notation, dependence of $j$ from $i$ is suppressed. Thus, the individual unit of observation is an operational site, and an event history instance has the form
\begin{equation*}
    (t_{i1},\dots,t_{in_{i}}:i=1,\dots,n)
\end{equation*}
where event times are grouped by operational site. Individual-specific historical information $H_{it}$, for $i=1,\dots,n$, can capture the observed heterogeneity in the event times of different operational sites. Furthermore, we introduce an individual-specific, potentially, time-varying, random effect $\zeta_{it}$ to induce the correlation structure between individuals. Hence, we can understand an operational site $i$ as sending $n_{i}$ requisitions, over a period of interest $(0,\tau]$, at times $0<t_{i1}<\cdots<t_{in_{i}}\le\tau$ with intensity $\lambda_{i}(\,\cdot\,|\,H_{it},\zeta_{it})$.

\subsubsection{Requisitions generation model}\label{sec:212}

The previous section describes an intensity model of the event times at which requisition generations are triggered. This section takes the next step to propose a mechanism of how a requisition's line items are filled. In more detail, the procedure of filling a requisition is seen as a sequential decision mechanism of the joint decision about whether to keep adding line items or stop, which product to introduce, and at what quantity. We explain the logic of the proposed mechanism in a simple example and provide a formal expression.
\begin{example}
    Suppose an agent at some operational site $i$ is preparing the $j$-th requisition to be sent from that particular site to procurement. Let $e_{k-1}\doteq e_{ij,k-1}$ represent the vector of product ids and relative quantities already included in the requisition at step $k$, $a_{k}^{(0)}\doteq a_{ijk}^{(0)}\in\{0,1\}$ is the decision about whether to introduce a $k$-th line item or not, $a_{k}^{(1)}\doteq a_{ijk}^{(1)}$ is the $k$-th product id, and $q_{k}\doteq q_{ijk}$ the relative quantity in the case that $a_{k}^{(0)}=1$; if $a_{k}^{(0)}=0$, the agent stops adding line items to the $j$-th requisition. Initially, it is clear that $e_{0}=\emptyset$ and $a_{1}^{(0)}=1$ since no item has been added yet. Next, the agent chooses the first line item and, hence, $e_{1}\doteq(a_{1}^{(1)},q_{1})$. Now, if $a_{2}^{(0)}=1$, then $e_{2}\doteq(a_{1}^{(1)},q_{1},a_{2}^{(1)},q_{2})$ and the process continues, for instance, until $K\doteq K_{ij}$ line items are added and the final requisition is $e_{ij}\doteq(a_{1},q_{1},\dots,a_{K},q_{K})$, where, clearly, $a_{k}\doteq(1,a_{k}^{(1)})$.
\end{example}
The logic that is illustrated in the example above can be formalised as follows. Consider the probability density function of an instance $(q_{ijk},a_{ijk})$ factorised as
\begin{equation}\label{eq:dem2seqdec}
    p(q_{ijk},a_{ijk}\,|\,e_{ij,k-1},H_{it_{ij}})=p(a_{ijk}^{(0)}\,|\,e_{ij,k-1},H_{it_{ij}})\,p(q_{ijk},a_{ijk}^{(1)}\,|\,e_{ij,k-1},H_{it_{ij}})^{a_{ijk}^{(0)}}
\end{equation}
where $e_{ijk}\doteq\big(t_{ij},(a_{ij1},q_{ij1}),\dots,(a_{ijk},q_{ijk})\big)$ denotes part of a requisition with $k\le K_{ij}$, $H_{it_{ij}}$ is the historical process explained in the previous section, and $(a_{ij},q_{ij})\doteq((a_{ijk},q_{ijk}):k=1,\dots,K_{ij})$ line items relative to $e_{ij}$. Hence, a realisation of the form
\begin{equation*}
    (e_{i1},\dots,e_{in_{i}}:i=1,\dots,n)
\end{equation*}
is sampled from
\begin{equation}\label{eq:dem2mod}
    \prod_{i=1}^{n}\prod_{j=1}^{n_{i}}\prod_{k=1}^{K_{ij}}p(q_{ijk},a_{ijk}\,|\,e_{ij,k-1},H_{it_{ij}},\zeta_{it_{ij}})
\end{equation}
where $H_{it_{ij}}$ contains historical data to adjust for observed heterogeneity in the content of requisitions, while the random vector $\zeta_{it_{ij}}$ controls for correlation patterns in the structure of requisitions generated at different operational sites, e.g., products often requested together.

Conversion of the demand from its own time scale, guided by the time-to-request model, to calendar days, or, in general, to a discrete time scale, is important when the decisions are carried out at discrete stages such as calendar days. Requisitions can be grouped so that demand occurs on daily basis
\begin{equation*}
    b_{i\ell}\doteq(e_{ij}:24(\ell-1)\le t_{ij}<24\ell:j=1,\dots,n_{i})
\end{equation*}
so that $b_{\ell}\doteq(b_{i\ell}:i=1,\dots,n)$. In this study, the aim is to simulate real world procurement operations and, thus, it is natural to think that decision-making takes place at a discrete calendar day based time scale.

\subsection{Decision model}\label{sec:22}

In this section, we describe a model of the operational (purchasing) decisions, i.e., when to order and supplier selection, in procurement. Consider a $\tau$-step realisation from the system given by
\begin{equation*}
    (b_{1},d_{1},y_{1}),\dots,(b_{\tau},d_{\tau},y_{\tau})
\end{equation*}
with probability density
\begin{equation}\label{eq:joint3}
    p(b_{1},d_{1},y_{1},\dots,b_{\tau},d_{\tau},y_{\tau}\,|\,\overline{X}_{\tau})=\prod_{\ell=1}^{\tau}p(b_{\ell}\,|\,b_{\ell}^{-},\overline{X}_{\ell}^{(b)})\,p(d_{\ell}\,|\,d_{\ell}^{-},\overline{X}_{\ell}^{(d)})\,p(y_{\ell}\,|\,y_{\ell}^{-},\overline{X}_{\ell}^{(y)})
\end{equation}
where $\overline{X}_{\ell}\doteq(\overline{X}_{\ell}^{(b)},\overline{X}_{\ell}^{(d)},\overline{X}_{\ell}^{(y)})$ is exogenous covariate history, $y_{\ell}^{-}\doteq(\overline{b}_{\ell},\overline{d}_{\ell},\overline{y}_{\ell-1})$, $d_{\ell}^{-}\doteq(\overline{b}_{\ell},\overline{d}_{\ell-1},\overline{y}_{\ell-1})$, and $b_{\ell}^{-}\doteq(\overline{b}_{\ell-1},\overline{y}_{\ell-1})$. The following assumption eliminates unnecessary layers of complexity. For instance, suppose the aim is to develop an optimal policy for processing the unresolved and new requests. The approach could be to model the daily buyer-supplier interaction by harvesting bits of information from placing many small orders on the same day. This might sound appealing, but goes far beyond what is realistic and to the best of our knowledge has never been attempted.
\begin{assumption}\label{asm:simp}
    Every day $\ell$, all line items in the array of unresolved requests as well as the ones arriving during the day are considered only once, in the chronological order they arrived, and all orders are placed at once at the end of the day.
\end{assumption}

A direct concomitant of the above assumption is that outcomes or outcome beliefs about the orders placed on a given day are available the next day. Let
\begin{equation*}
    \mathcal{I}(\ell)\doteq\{(l,m)\in\{(1,1),\dots,(1,M_{1}),\dots,(\ell,1),\dots,(\ell,M_{\ell})\}:(d_{lm}^{(0)}=0)\vee(l=\ell)\}
\end{equation*}
index the set of unresolved requests on day $\ell$. For a single instance $\ell$, we may isolate the decision mechanism
\begin{align}\label{eq:decmec}
    p(d_{\ell}\,|\,d_{\ell}^{-},\overline{X}_{\ell}^{(d)})&=\prod_{(l,m)\in\mathcal{I}(\ell)}p(d_{lm}\,|\,d_{lm}^{-}(\ell),\overline{X}_{\ell}^{(d)})\nonumber\\
    &=\prod_{(l,m)\in\mathcal{I}(\ell)}p(d_{lm}^{(0)},d_{lm}^{(1)}\,|\,d_{lm}^{-}(\ell),\overline{X}_{\ell}^{(d)})\\
    &=\prod_{(l,m)\in\mathcal{I}(\ell)}p(d_{lm}^{(0)}\,|\,d_{lm}^{-}(\ell),\overline{X}_{\ell}^{(d)})\,p(d_{lm}^{(1)}\,|\,d_{lm}^{-}(\ell),\overline{X}_{\ell}^{(d)})^{d_{lm}^{(0)}}\nonumber
\end{align}
where $d_{lm}^{-}(\ell)\doteq\big(d_{\ell}^{-},d_{l'm'}:(l',m')\in\mathcal{I}(\ell):(l'<l)\vee[(l'=l)\wedge(m'<m)]\big)$, under Assumption \ref{asm:simp}, contains information about line items processed at day $\ell$ before line item $(l,m)$. Clearly, if $d_{lm}^{(0)}=0$, then $d_{lm}^{(1)}=\emptyset$, which means it is not observed and the line item index $(l,m)$ should enter the set $\mathcal{I}(\ell+1)$, i.e., the next day's list of unresolved requests (including the ones generated on day $\ell+1$).

\subsection{Outcome model}

Consider again Equation~\eqref{eq:joint3} and take an instance of the outcome variable $y_{\ell}$. We can write its probability density as
\begin{equation}\label{eq:respmec}
    p(y_{\ell}\,|\,y_{\ell}^{-},\overline{X}_{\ell}^{(y)})=\prod_{(l,m)\in\mathcal{I}(\ell)}p(y_{lm}\,|\,y_{\ell}^{-},\overline{X}_{\ell}^{(y)})^{d_{lm}^{(0)}}
\end{equation}
so that the generating mechanism of the first stage decision, $d_{lm}^{(0)}$, on whether to place an order at day $\ell$ for line item $(l,m)$, acts again as a missingness mechanism for the supplier selection and response vector. Assumption \ref{asm:simp} does not allow the values $y_{lm}$ to be sampled before the next day of the corresponding supplier selection $d_{lm}^{(1)}$.

We emphasise that the outcome variable $y_{\ell}$ contains the actual cost of line items in the orders placed at day $\ell$, but information regarding the (delivery) lead time and quality are only the supplier's estimates and can be different than the actual ones. We therefore introduce the following assumption.
\begin{assumption}[Supplier honesty]
    Suppliers provide consistent estimates of the expected lead time and quality of an order.
\end{assumption}
The above assumption alleviates the need to develop a separate model of the actual performance of suppliers as compared to the one disclosed in their estimates as it asserts that significant divergence between the supplier-provided outcome estimates and the actual outcomes is not due to systemic behavior but, rather, could only be caused by exogenous effects. Supplier honesty is a mild requirement as it is expected that any procurement organisation carries out a careful supplier on-boarding process prior to allocating business.

\subsection{Discrete-event simulation logic}

The theory of discrete-event simulation (DES) modelling has been described extensively by \cite{law2000des} who suggests the collection of state variables, a simulation clock, and an event list as the necessary components of any DES model. The simulation clock is a variable in the simulation model which registers the current value of simulated time, and the event list contains the next time each type of event occurs. In this section, we provide all the details of the proposed DES logic for the procurement operations system.

The evolution of a procurement system is the sequence of transitions between states at discrete points in time; the state of the system at time $t$ is determined by historical demand, decisions, and outcomes. An operational site $i$ generates its $j$-th requisition $e_{ij}$ at time $t_{ij}$ adding an increment to the total count of unresolved requisitions $|\mathcal{I}(\ell)|$, before the upcoming decision point $t=\ell$, with $\ell\in\{1,2,\dots\}$. By the end of day $\ell$ a number of requisitions has been processed and orders are placed at the end of day $\ell$ to generate the decision and outcome vectors, $d_{\ell}$ and $y_{\ell}$, respectively.

Five event types, that is, discrete events that take place in procurement operations are summarised in Table~\ref{tab:desproc-ev}. The first three event types are the crucial ones, whereas the latter two are included as potentially useful for case-specific designs, more precisely, the event types 1 and 2 drive the evolution, whereas event type 3 signals the termination of the simulation. Event type 4 is an evaluation routine that simply produces summary measures to monitor the performance of different suppliers every quarter of a calendar year, and event type 5 represents the time an order has been delivered.

\begin{table}
\centering
\caption{Events taking place in procurement operations.}
{\begin{tabular}{lc} \toprule
        Event description & Event type \\
        \midrule
        The generation of a requisition & 1 \\
        Decision point every calendar day & 2 \\
        Termination of the simulation after $\tau$ calendar days & 3 \\
        Supplier evaluation (honesty assessment) every 90 days & 4 \\ % honesty assessment
        Order delivery & 5 \\
        \bottomrule
\end{tabular}}
\label{tab:desproc-ev}
\end{table}

\begin{figure}[ht]
    \centering
    \includegraphics[width=0.8\textwidth]{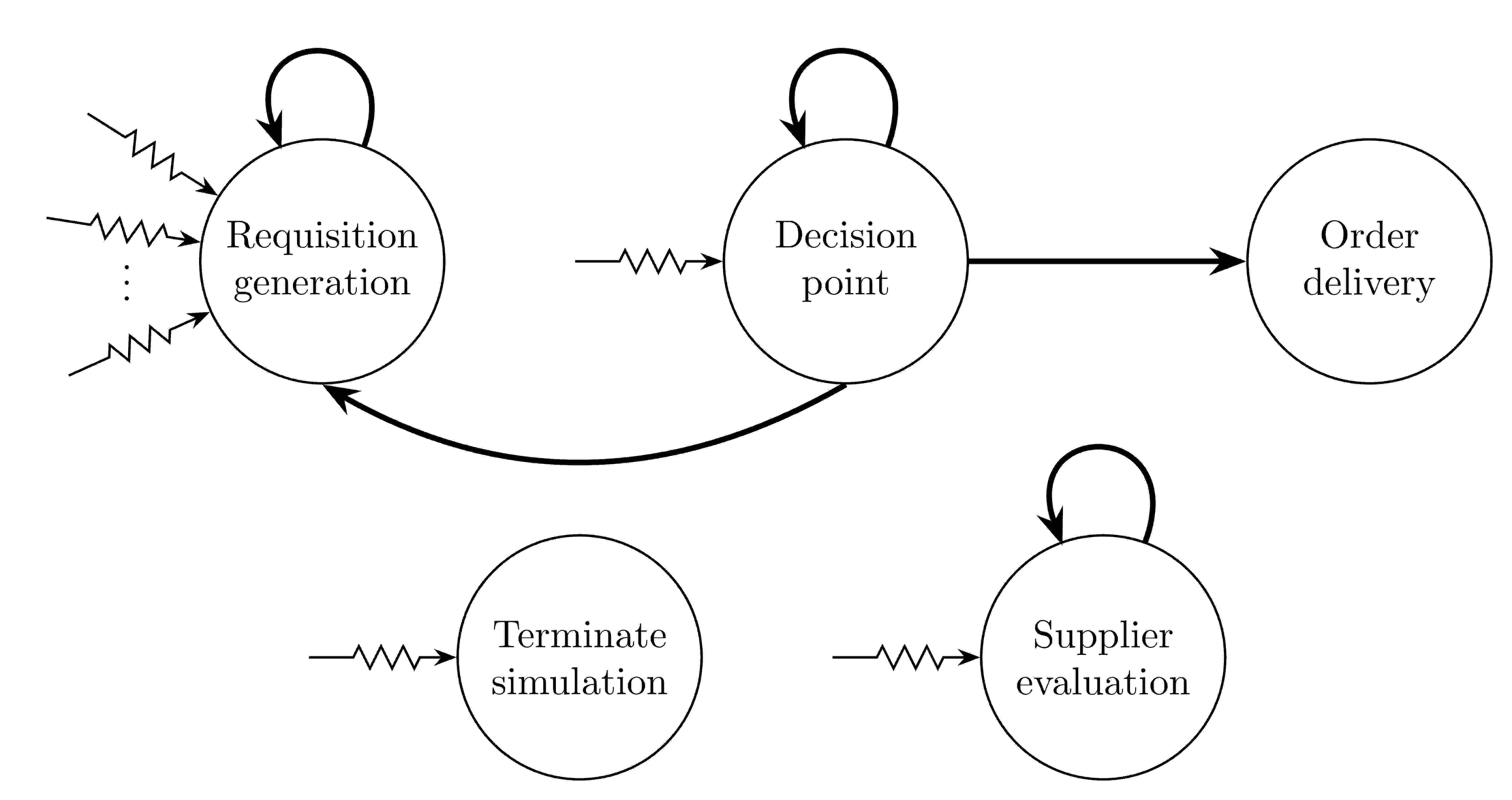}
    \caption{Event graph outline of the proposed procurement simulation model. A single jagged arrow pointing to an event means that an event of the sort is scheduled when the simulation is initiated, whereas multiple jagged arrows mean that an event is scheduled initially for each operational site. A thick arrow indicates that the event type at the start schedules the next event of the type at the tip of the arrow.}
	\label{fig:eventgraph}
\end{figure}

A more technical presentation of the proposed DES model can be guided by the event graph \citep{law2000des}. The purpose of an event graph is to illustrate the scheduling of the event types. In our DES model, the first requisition generation for each operational site, the first decision point, the first supplier evaluation, and the termination of the simulation are all scheduled at initialisation. A requisition generation, a decision point, or a supplier evaluation event schedule the time of the next event of the same sort. Moreover, a decision point event can schedule the time of the next order delivery event too as well as the next requisition generation event. Figure~\ref{fig:eventgraph} depicts the event graph.

\section{Proposed simulation model}\label{sec:propmod}

In this section, we provide detailed specifications and algorithmic implementations of the different components of the probabilistic model formalised in the previous section. Each section contains a mathematical expression, specifications, and a simulation algorithm of the corresponding model.

\subsection{Generation of request times}

The first stage of demand generation involves an intensity model to trigger the generation of requisitions. Recall that products are grouped into categories which are known a priori so that a demand model is specific to a category of products. Consider the intensity function having the multiplicative form, parameterised by the Euclidean vector of, potentially, time-varying parameters $\beta_{t}$,
\begin{equation}\label{eq:multmod}
    \lambda_{i}(t\,|\,H_{it},\zeta_{i};\beta_{t})\doteq\zeta_{i}^{(1)}\lambda_{i0}(t)\exp(\widetilde{H}_{it}^{\top}\beta_{it})
\end{equation}
where $\lambda_{i0}(t)$ is a positive-valued function called the baseline intensity, $\widetilde{H}_{it}$ are features, i.e., main effects, higher order interactions, or basis expansions, constructed by applying transformations to $H_{it}$, and correlation patterns between operational sites are introduced through unobserved effects $\zeta_{i}^{(1)}$, for $i=1,\dots,n$, assumed to be i.i.d. draws from the same distribution with $E\zeta_{i}^{(1)}=1$.

\subsubsection*{Specification}

Features can be constructed from historical data $H_{t}$ by applying a mapping
\begin{equation*}
    H_{t}\mapsto f_{b,t}^{(1)}(H_{t})\doteq\widetilde{H}_{t}
\end{equation*}
to obtain a vector of main effects as well as interactions and basis expansions.

We propose a total of six main effects plus baseline characteristics $\widetilde{H}_{0}$ that influence the time a requisition is generated, see Table~\ref{tab:tteft}. Efficient computation of such features requires operational site-specific records of event times, lead time, and quality to be stored as the simulation is performed. We can then simply extract the latest recorded delivery day, for every operational site $i$, and subtract from the present day to obtain the first main effect, while the second would equal the last quality record for each $i$. Clearly, these two effects do not vary within a given day. However, computation of the third feature requires the information collected during day $\ell$. Business policy shocks and market disruptions are represented with sinc functions as we expect they would cause spikes in the demand over narrow intervals, whereas seasonal trends are captured with calendar time harmonics. There is no computational burden associated with the exogenous features since they are computed prior to the simulation run. % The mapping
% \begin{equation*}
%     \overline{X}_{\ell}^{(b)}\mapsto f_{x,\ell}^{(b,1)}(\overline{X}_{\ell}^{(b)})\doteq\left(\widetilde{X}_{\ell}^{(b,11)\top},\widetilde{X}_{\ell}^{(b,12)\top},\widetilde{X}_{\ell}^{(b,13)\top}\right)^{\top}\doteq\widetilde{X}_{\ell}^{(b,1)}
% \end{equation*}
% simply returns the $\ell$-th value of each pre-computed array. Notice that the proposed exogenous covariates are evaluated on a daily resolution, and, therefore, $\widetilde{X}_{\ell}^{(b,1)}=\widetilde{X}_{t}^{(b,1)}$ for all $24\ell\le t<24(\ell+1)$.

\begin{table}
\centering
\caption{Main effects in the intensity model constructed from historical data.}
{\begin{tabular}{rlll}
        \toprule
        & Short description & Symbol & Level\\
        \midrule
        1 & Days since latest delivery & $\widetilde{H}_{t}^{(1)}$ & Category \\
        2 & Quality of the latest delivery & $\widetilde{H}_{t}^{(2)}$ & Category \\
        3 & Number of requisitions sent over the past year & $\widetilde{H}_{t}^{(3)}$ & Category \\
        4 & Seasonal trends & $\widetilde{H}_{t}^{(4)}$ & Category \\
        5 & Business policy shock & $\widetilde{H}_{t}^{(5)}$ & Category \\
        6 & Market disruption & $\widetilde{H}_{t}^{(6)}$ & Category \\
        7 & Baseline information & $\widetilde{H}_{0}$ & Operational site \\
        \bottomrule
    \end{tabular}}
% \tabnote{\textsuperscript{a}This footnote shows how to include
%  footnotes to a table if required.}
\label{tab:tteft}
\end{table}

Eventually, the unobserved effects $\zeta_{i}^{(1)}$ are taken to be Gamma distributed, for $i=1,\dots,n$, with mean equal to $1$, and we suggest each baseline intensity $\lambda_{i0}(t)$ be either parameterised as a Weibull distribution or as a piecewise-constant function \citep{cook2007a}.

\subsubsection*{Simulation}

To simulate a recurrent event process with intensity function \eqref{eq:multmod} and time-varying covariates, we propose an extension of the thinning algorithm of \cite{lewis1979a} to accommodate observed heterogeneity that is captured by the time-varying covariate vector $\widetilde{H}_{t}$. Algorithm~\ref{algo:timetoreq} simulates a single event time $t_{ij}$ when a requisition is to be generated. The algorithm requires the norm of $\widetilde{H}_{t}$ to be bounded. The function $\lambda^{*}$ is an upper bound on the intensity function.
\begin{table}
\small
% \caption{Meta-Coal Algorithm}
{\begin{algorithm}[H]
    \label{algo:timetoreq}
    % \SetKwInOut{Input}{Input}
    % \SetKwInOut{Output}{Output}
    % \SetAlgorithmName{Algorithm}{}
    \KwIn{$t_{i,j-1}$, $\widetilde{H}_{it}$, $\lambda_{i}$, $\lambda_{i}^{*}$, $\zeta_{i}^{(1)}$, $\tau$}
    \KwOut{event time $t_{ij}$}
    Set $t_{ij}$ to the previous event time $t_{i,j-1}$\;
    \While{$t_{ij}\le\tau$}{
        Draw $v$ from Uniform$(0,1)$\;
        \If {$v<\zeta_{i}^{(1)}\dfrac{\lambda_{i}}{\lambda_{i}^{*}}(t_{ij},\widetilde{H}_{it})$}{$t_{ij}\leftarrow t_{ij}-\log(v)/\lambda_{i}^{*}(t_{ij})$\;\textbf{break}\;}{}
    }
-    \caption{Time-to-request mechanism.}
\end{algorithm}}
\end{table}

\subsection{Generation of requisitions}\label{sec:32}

The second component of the demand model is concerned with the generation of requisitions. Consider a requisition instance generated, according to \eqref{eq:dem2mod}, by
\begin{equation*}
    \prod_{k=1}^{K}p(q_{k},a_{k}\,|\,e_{k-1},H_{it},\zeta_{i})
\end{equation*}
where we dispense with the subscripts $i,j$, for brevity. The value of $K$ is implicitly set by a sequential two stage decision $a_{k}\doteq(a_{k}^{(0)},a_{k}^{(1)})$; $a_{k}^{(0)}\in\{0,1\}$ indicates whether to add another line item in the current requisition or stop, and $a_{k}^{(1)}\in\mathcal{A}$ the product identifier. By $\mathcal{A}$ we denote the finite set of all discrete product identifiers for a given category. Hence, we can factorise the mechanism further as
\begin{equation*}
\begin{aligned}
    p(q_{k},a_{k}\,|\,e_{k-1},H_{it},\zeta_{i}^{(2)})&=p(a_{k}^{(0)}\,|\,e_{k-1},H_{it})\,p(a_{k}^{(1)}\,|\,e_{k-1},H_{it})^{a_{k}^{(0)}}\\
    &\hspace{15pt}\times p(q_{k}\,|\,a_{k}^{(1)},e_{k-1},H_{it},\zeta_{i})^{a_{k}^{(0)}}
\end{aligned}
\end{equation*}
% where the historical data $H_{it}$ are used to construct a set of features for the first stage decision, second stage decision, and quantity models.
A natural modelling choice for the first stage decision-to-stop variable $a_{k}^{(0)}$ is the logistic regression model with propensity score given by
\begin{equation}\label{eq:r1st}
    \text{logit}\left(U_{k}^{(e,0)}\right)\doteq\widetilde{e}_{k}^{(0)\top}\gamma_{\ell}^{(0)}
\end{equation}
where $(e_{k-1},H_{it})\mapsto f_{e,tk}^{(0)}(e_{k-1},H_{it})\doteq\widetilde{e}_{k}^{(0)}$ are features derived from $(e_{k-1},H_{it})$, $\gamma_{\ell}^{(0)}$ an Euclidean parameter vector.

Another intuitive modelling option is to let the second stage product selection variable $a_{k}^{(1)}$ follow a mixed multinomial distribution with utility model
\begin{equation}\label{eq:crewutil}
    U_{k}^{(e,1)}(a;w)\doteq a_{k}^{(0)}\left(\widetilde{e}_{k}^{(1)}(a)^{\top}\gamma_{\ell}^{(1)}+G_{a}^{\top}w\right)
\end{equation}
where $(a,e_{k-1},H_{it})\mapsto f_{e,tk}^{(1)}(a,e_{k-1},H_{it})\doteq\widetilde{e}_{k}^{(1)}(a)$ are features specific to product index $a$ computed from $(a,e_{k-1},H_{it})$, respectively, $\gamma_{\ell}^{(1)}$ a Euclidean parameter vector, $w$ are random intercepts aiming to capture co-occurrence patterns between products, and $G$ is a design matrix defining the correlation groups; $G_{a}$ denotes the column of $G$ that corresponds to product index $a$. The propensity score of selecting product $a$ is given as
\begin{equation}\label{eq:itemprop}
    p(a\,|\,e_{k-1},H_{it};\gamma_{\ell}^{(1)},w)\doteq\frac{\exp\left(U_{k}^{(e,1)}(a;w)\right)}{\sum_{a'\in\mathcal{A}}\exp\left(U_{k}^{(e,1)}(a';w)\right)}
\end{equation}

Eventually, suppose $Q_{ijk}$ takes non-negative integer values, for $k=1,\dots,K_{ij}$, and the quantities of different products and services in the same requisition are independent given the historical data, that is, $Q_{ijk}\idp Q_{ijk'},A_{ijk'}\,|\,A_{ijk},H_{it}$ for all $k'\ne k$. In this case, the quantity values can be sampled from a mechanism of the sort
\begin{equation*}
    p(q\,|\,a^{(1)},e_{k-1},H_{it},\zeta_{i})\doteq\prod_{k=1}^{K}p(q_{k}\,|\,a_{k}^{(1)},H_{it},\zeta_{i})
\end{equation*}
Let $(a_{k}^{(1)},H_{it})\mapsto f_{e,tk}^{(2)}(a_{k}^{(1)},H_{it})\doteq\widetilde{e}_{k}^{(2)}(a_{k}^{(1)})$ denote summary features constructed from the subset $H_{it}$ relevant for product index $a_{k}^{(1)}$, $\varpi$ a function associated with the mapping $\widetilde{e}_{k}^{(2)}(a_{k}^{(1)})\mapsto\varpi\left(\widetilde{e}_{k}^{(2)}(a_{k}^{(1)})\right)\doteq\varpi_{k}$, and $\zeta_{i}^{(2)}$ a random effect that induces correlation in the quantity mechanism between different operational sites. In this case, the quantity values can be sampled from a random distribution with density function
\begin{equation}
\begin{aligned}
    p(q\,|\,a^{(1)},e_{k-1},H_{it},\zeta_{i})=\prod_{k=1}^{K}&\frac{\left(\zeta_{i}^{(2)}\varpi_{k}\right)^{q_{k}}}{q_{k}!}\exp\left(-\zeta_{i}^{(2)}\varpi_{k}\right)
\end{aligned}
\end{equation}

\subsubsection*{Specification}

The requisition generation model consists of three sub-models, presented above, each of which may require its own set of features and parameter specifications. Table~\ref{tab:d2specs} summarises the main effects to be considered in each model; the row numbering convention is such that the first digit indicates a sub-model, while the second enumerates the effects. The letter `x' is used in place of the first digit to indicate that the relative effect may be used by more than one model.

Two covariates are suggested to bring an effect on the decision-to-stop variable. Computation of $\widetilde{e}_{k}^{(01)}$ requires storing the current number of line items in a requisition, while $\widetilde{e}_{k}^{(02)}$ needs an array of quantities and product identifiers of the line items currently in the requisition as well as unit cost estimates of the relative products, e.g., a rolling average of the most recent unit cost realisations. The main effects $\widetilde{e}_{k}^{(11)},\dots,\widetilde{e}_{k}^{(15)}$ are used in Expression~\eqref{eq:crewutil} of the product utility model. Computation of $\widetilde{e}_{k}^{(11)}$ requires that the number of requests for each product and operational site combination over the past year is recorded. For $\widetilde{e}_{k}^{(12)}$ a rolling sum of the annual quantity of each product delivered at each operational site must be maintained. Effect $\widetilde{e}_{k}^{(13)}$ can be computed by keeping track of the latest day each product was requested from each operational site. The value can be simply computed by extracting the relative record and subtracting it from the present day. Similarly, tracing the day of the latest delivery for each product and operational site pair would enable the computation of $\widetilde{e}_{k}^{(14)}$ by subtracting the present day from the desired record. Eventually, records of the observed quality of the latest delivery on all product and operational site pairs can be kept to be able to derive $\widetilde{e}_{k}^{(15)}$ by extracting the latest record for the desired combination. Computation of main effects $\widetilde{e}_{k}^{(21)}$ and $\widetilde{e}_{k}^{(22)}$ which could influence the quantity mechanism may require records of the total quantity requested and ordered for each pair of product and operational site over the past year. Additional exogenous effects can be considered in any model to incorporate the effect of seasonal trends, business policies, market disruptions, or baseline characteristics of products and operational sites on the corresponding decision variable. Finally, it is worth noting that Table~\ref{tab:d2specs} illustrates only main effects but higher order interactions and basis expansions could also enter the modelling expression.

\begin{table}
\centering
\caption{Main effects in the requisition generation model constructed from historical data.}
{\begin{tabular}{rlll}
        \toprule
        & Short description & Symbol & Level\\
        \midrule
        01 & Number of line items in requisition & $\widetilde{e}_{k}^{(01)}$ & Requisition \\
        02 & Cost estimate of requisition & $\widetilde{e}_{k}^{(02)}$ & Requisition \\
        11 & Number of requests over the past year & $\widetilde{e}_{k}^{(11)}$ & Product \\
        12 & Quantity delivered over the past year & $\widetilde{e}_{k}^{(12)}$ & Product \\
        13 & Number of days since last request & $\widetilde{e}_{k}^{(13)}$ & Product \\
        14 & Number of days since last delivery & $\widetilde{e}_{k}^{(14)}$ & Product \\
        15 & Quality of last delivery & $\widetilde{e}_{k}^{(15)}$ & Product \\
        21 & Quantity requested over the past year & $\widetilde{e}_{k}^{(21)}$ & Product \\
        22 & Quantity delivered over the past year & $\widetilde{e}_{k}^{(22)}$ & Product \\
        x1 & Seasonal trends & $\widetilde{e}_{k}^{(03)},\widetilde{e}_{k}^{(16)},\widetilde{e}_{k}^{(23)}$ & Product \\
        x2 & Business policy shock & $\widetilde{e}_{k}^{(04)},\widetilde{e}_{k}^{(17)},\widetilde{e}_{k}^{(24)}$ & Product \\
        x3 & Market disruption & $\widetilde{e}_{k}^{(05)},\widetilde{e}_{k}^{(18)},\widetilde{e}_{k}^{(25)}$ & Product \\
        x4 & Baseline information & $\widetilde{e}_{k}^{(06)}$ & Operational site \\
        x5 & Baseline information & $\widetilde{e}_{k}^{(19)},\widetilde{e}_{k}^{(26)}$ & Product \\
        \bottomrule
    \end{tabular}}
% \tabnote{\textsuperscript{a}This footnote shows how to include
%  footnotes to a table if required.}
\label{tab:d2specs}
\end{table}

The proposed choices for the functional form of $\varpi$ and the distribution of the random effect $\zeta_{i}^{(2)}$ are to use a simple additive model structure parameterised with a Euclidean vector $\gamma_{\ell}^{(2)}$ for the former and a Gamma distribution with mean equal to 1 for the latter.

\subsubsection*{Simulation}

The logic of requisition generation in summarised in Algorithm~\ref{algo:reqgen}; notice that the elements $\mathcal{A}$, $G$, $\Sigma_{\mathcal{A}}$, $\gamma_{\ell}^{(0)}$, $\gamma_{\ell}^{(1)}$, $\varpi$, $\phi_{2}$, and transformations $f_{e,t}^{(0)}$, $f_{e,t}^{(1)}$, $f_{e,t}^{(2)}$, where $f_{e,t}^{(\cdot)}$ contains all $f_{e,tk}^{(\cdot)}$, are configuration parameters and must be provided as input along with historical data $H_{it}$. The symbol $\mathcal{N}(\,\cdot\,,\cdot\,)$ is used to denote a normal distribution.

% \begin{enumerate}
%     \item Initialisation: set $e$ to the empty set, $\bar{\mathcal{A}}$ to $\mathcal{A}$, $a_{0}$ and $k$ to 1, draw $w$ from $\mathcal{N}(0,\Sigma_{\mathcal{A}})$ and $\zeta^{(2)}$ from Gamma$(\phi_{2}^{-1})$
%     \item for each $a\in\bar{\mathcal{A}}$
%     \begin{enumerate}
%         % \item draw $\epsilon^{(e,1)}(a)$ from $\mathcal{N}(0,\sigma_{e,1}^{2})$
%         \item set $U^{(e,1)}(a)$ to $f_{e,tk}^{(1)}(a,e_{k-1},H_{it})^{\top}\gamma_{\ell}^{(1)}+G_{a}^{\top}w_{a}$ % +\epsilon^{(e,1)}(a)$
%         \item set $p^{(e,1)}(a)$ to $\exp\left(U^{(e,1)}(a)\right)\Big/\sum_{a'\in\bar{\mathcal{A}}}\exp\left(U^{(e,1)}(a')\right)$
%     \end{enumerate}
%     \item draw $a_{k}^{(1)}$ from a multinomial distribution over $\bar{\mathcal{A}}$ with propensity vector $p^{(e,1)}$ and include it in $e$ and remove $a_{k}^{(1)}$ from $\bar{\mathcal{A}}$
%     \item sample the quantity $q_{k}$ of $a_{k}^{(1)}$ from a negative binomial distribution with intensity $\zeta^{(2)}\varpi\left(f_{e,tk}^{(2)}(a_{k}^{(1)},H_{it})\right)$
%     % \item draw $\epsilon^{(e,0)}$ from $\mathcal{N}(0,\sigma_{e,0}^{2})$ and set $U^{(e,0)}$ to $\text{expit}\left(\widetilde{e}_{k}^{(0)\top}\gamma_{\ell}^{(0)}+\epsilon_{k}^{(e,0)}\right)$
%     \item set $U^{(e,0)}$ to $\text{expit}\left(\widetilde{e}_{k}^{(0)\top}\gamma_{\ell}^{(0)}\right)$
%     \item draw $a^{(0)}$ from $\text{Bernoulli}\left(U^{(e,0)}\right)$, if $a^{(0)}=1$ and $|\bar{\mathcal{A}}|>0$ set $k$ to $k+1$, go to (2)
% \end{enumerate}

\begin{table}
\small
% \caption{Meta-Coal Algorithm}
{
\begin{algorithm}[H]
    \label{algo:reqgen}
    % \SetKwInOut{Input}{Input}
    % \SetKwInOut{Output}{Output}
    % \SetAlgorithmName{Algorithm}{}
    \KwIn{$\mathcal{A}$, $G$, $\Sigma_{\mathcal{A}}$, $\gamma_{\ell}^{(0)}$, $\gamma_{\ell}^{(1)}$, $\varpi$, $\phi_{2}$, $f_{e,t}^{(0)}$, $f_{e,t}^{(1)}$, $f_{e,t}^{(2)}$, $H_{it}$}
    \KwOut{requisition $e$}
    \textbf{Initialisation:} $e\leftarrow\emptyset$, $\bar{\mathcal{A}}\leftarrow\mathcal{A}$, $a_{0}\leftarrow k\leftarrow1$, draw $w$ from $\mathcal{N}(0,\Sigma_{\mathcal{A}})$ and draw $\zeta^{(2)}$ from Gamma$(\phi_{2}^{-1})$\;
    \ForEach{$a\in\bar{\mathcal{A}}$}{
    $U^{(e,1)}(a)\leftarrow f_{e,tk}^{(1)}(a,e_{k-1},H_{it})^{\top}\gamma_{\ell}^{(1)}+G_{a}^{\top}w$\;
    $p^{(e,1)}(a)\leftarrow\exp\left(U^{(e,1)}(a)\right)\Big/\sum_{a'\in\bar{\mathcal{A}}}\exp\left(U^{(e,1)}(a')\right)$\;
    Draw $a_{k}^{(1)}$ from a multinomial distribution over $\bar{\mathcal{A}}$ with propensity vector $p^{(e,1)}$, push it in $e$, and remove $a_{k}^{(1)}$ from $\bar{\mathcal{A}}$\;
    Sample the quantity $q_{k}$ of $a_{k}^{(1)}$ from a negative binomial distribution with intensity $\zeta^{(2)}\varpi\left(f_{e,tk}^{(2)}(a_{k}^{(1)},H_{it})\right)$\;
    $U^{(e,0)}\leftarrow\text{expit}\left(\widetilde{e}_{k}^{(0)\top}\gamma_{\ell}^{(0)}\right)$\;
    Draw $a^{(0)}$ from $\text{Bernoulli}\left(U^{(e,0)}\right)$\;
    \eIf {$a^{(0)}=0$ or $|\bar{\mathcal{A}}|\le0$}{\textbf{break}\;}{$k\leftarrow k+1$\;}}
\caption{Requisition generation mechanism.}
\end{algorithm}}
\end{table}

\subsection{Decision and outcome models}\label{sec:33}

\subsubsection{Decision model}

The decision mechanism is given in Equation~\eqref{eq:decmec} where we show that on a given day indexed by $\ell$ decisions are made for each line item in the array of unresolved requests. More precisely, a decision is performed by some buyer and involves two stages, first, whether to order the line item or not, and, second, the selection of a supplier. The formula is repeated below:
\begin{equation*}
    \prod_{(l,m)\in\mathcal{I}(\ell)}p(d_{lm}^{(0)}\,|\,d_{lm}^{-}(\ell),\overline{X}_{\ell}^{(d)})\,p(d_{lm}^{(1)}\,|\,d_{lm}^{-}(\ell),\overline{X}_{\ell}^{(d)})^{d_{lm}^{(0)}}
\end{equation*}
A natural choice of a model for the first stage decision variable $d_{lm}^{(0)}$ is a logistic regression with propensity score
\begin{equation}
    \text{logit}\left(U_{lm}^{(d,0)}\right)=\widetilde{d}_{lm}^{(0)\top}\theta_{\ell}^{(0)}
\end{equation}
where $(d_{lm}^{-}(\ell),\overline{X}_{\ell}^{(d)})\mapsto\left(f_{d,lm}^{(0)}(d_{lm}^{-}(\ell)),f_{x,\ell}^{(d,0)}(\overline{X}_{\ell}^{(d)})\right)\doteq\widetilde{d}_{lm}^{(0)}$ denotes the feature vector obtained by transforming the historical data. Notice the implicit dependence of an index $(l,m)$ from the calendar day index $\ell$ since for any index $(l,m)$ there exists a calendar index $\ell$ such that $(l,m)\in\mathcal{I}(\ell)$.

For supplier selection we propose the multinomial model with utility function
\begin{equation}\label{eq:suputil}
    U_{lm}^{(d,1)}(d)\doteq d_{lm}^{(0)}\left(\widetilde{d}_{lm}^{(1)\top}\theta_{\ell}^{(1)}\right)
\end{equation}
where $(d,d_{lm}^{-}(\ell),\overline{X}_{\ell}^{(d)})\mapsto\left(f_{d,lm}^{(1)}(d,d_{lm}^{-}(\ell)),f_{x,\ell}^{(d,1)}(d,\overline{X}_{\ell}^{(d)})\right)\doteq\widetilde{d}_{lm}^{(1)}(d)$ contains features specific to supplier option $d$, $\theta_{\ell}$ is a Euclidean parameter vector, and the propensity score for choosing supplier $d$ is given by
\begin{equation}
    p(d\,|\,d_{lm}^{-}(\ell),\overline{X}_{\ell}^{(d)};\theta_{\ell}^{(1)})=\frac{\exp(U_{lm}^{(d,1)}(d))}{\sum_{d'}\exp(U_{lm}^{(d,1)}(d'))}
\end{equation}
% On a given day $\ell$, the additive noise terms $\epsilon_{lm}^{(d,0)}$ and $\epsilon_{lm}^{(d,1)}$ are assumed to be i.i.d.

\subsubsection{Outcome model}

The outcome vector is modelled as a multivariate normal distribution of the sort
\begin{equation}
Y_{lm}=\widetilde{y}_{lm}^{\top}\delta_{\ell}+\epsilon_{lm}^{(y)}
\end{equation}
where the mapping $(y_{\ell}^{-},\overline{X}_{\ell}^{(y)})\mapsto f_{y,\ell}(y_{\ell}^{-},\overline{X}_{\ell}^{(y)})\doteq\widetilde{y}_{\ell}$ can be used to construct the features $\widetilde{y}_{lm}$, $\delta_{\ell}$ is a Euclidean vector of parameters, and $\epsilon_{lm}^{(y)}$ follows a 3-dimensional normal distribution centered around 0 and with covariance matrix $\Sigma$; notice that $\widetilde{y}_{\ell}\doteq(\widetilde{y}_{lm}:(l,m)\in\mathcal{I}(\ell))$.

\subsubsection*{Specification}

We elaborate on the efficient computation of the main effects that determine the decision-outcome dynamics; a detailed description of these effects is given in Table~\ref{tab:decspecs}. The effect $\widetilde{d}_{lm}^{(01)}$ can be computed using event time records for each requisition by extracting the relevant value and subtracting it from the current day. The value of $\widetilde{d}_{lm}^{(02)}$ can be obtained by continuously monitoring the number of requisitions in the list of unresolved requests. The mode of delivery and urgency status are disclosed in a requisition and can be retrieved with the event time of the corresponding requisition. The latest outcome records as well as an average outcome can be traced for each product and supplier combination. Computation of $\widetilde{d}_{lm}^{(13)}$ and $\widetilde{d}_{lm}^{(14)}$ requires maintaining an array with the total quantity of each product allocated to each supplier, so that the former can be obtained using the desired product and supplier combination, whereas the latter by summing over the quantity of products allocated to the supplier at question. Once more, exogenous effects would be seasonal trends, business policy shocks, or market disruptions, but also supplier or product baseline information.

\begin{table}
\centering
\caption{Main effects in the decision and outcome models constructed from historical data.}
{\begin{tabular}{rlll}
        \toprule
        & Short description & Symbol & Level\\
        \midrule
        01 & Number of days since generation & $\widetilde{d}_{lm}^{(01)}$ & Requisition \\
        02 & Number of unresolved requisition & $\widetilde{d}_{lm}^{(02)}$ & Total \\
        03 & Mode of delivery (local or stores) & $\widetilde{d}_{lm}^{(03)}$ & Requisition \\
        04 & Urgency status & $\widetilde{d}_{lm}^{(04)}$ & Requisition \\
        11 & Outcome of previous order from supplier & $\widetilde{d}_{lm}^{(11)}$ & Supplier and Product \\
        12 & Average outcome of supplier & $\widetilde{d}_{lm}^{(12)}$ & Supplier and Product \\
        13 & Product volume allocated to supplier & $\widetilde{d}_{lm}^{(13)}$ & Supplier and Product \\
        14 & Total volume allocated to supplier & $\widetilde{d}_{lm}^{(14)}$ & Supplier \\
        21 & Autoregressive terms & $\widetilde{y}_{lm}^{(1)}$ & Supplier or Product \\
        22 & Moving average terms  & $\widetilde{y}_{lm}^{(2)}$ & Supplier or Product \\
        23 & Recent volume ordered from supplier & $\widetilde{y}_{lm}^{(3)}$ & Supplier \\
        24 & Total volume ordered from supplier  & $\widetilde{y}_{lm}^{(4)}$ & Supplier \\
        x1 & Seasonal trends & $\widetilde{d}_{lm}^{(05)},\widetilde{d}_{lm}^{(15)},\widetilde{y}_{lm}^{(5)}$ & Supplier or Product \\
        x2 & Business policy shock & $\widetilde{d}_{lm}^{(06)},\widetilde{d}_{lm}^{(16)},\widetilde{y}_{lm}^{(6)}$ & Supplier or Product \\
        x3 & Market disruption & $\widetilde{d}_{lm}^{(07)},\widetilde{d}_{lm}^{(17)},\widetilde{y}_{lm}^{(7)}$ & Supplier or Product \\
        x4 & Baseline information & $\widetilde{d}_{lm}^{(08)},\widetilde{d}_{lm}^{(18)},\widetilde{y}_{lm}^{(8)}$ & Supplier or Product \\
        \bottomrule
    \end{tabular}}
% \tabnote{\textsuperscript{a}This footnote shows how to include
%  footnotes to a table if required.}
\label{tab:decspecs}
\end{table}

The main effects considered in the proposed outcome model are the recent and total quantity of products, also termed volume, allocated to the different suppliers. Computation of such effects requires maintaining volume information for each supplier as the simulation is running. Historical realisations of the outcome as well as noise terms are recorded and can be used in the outcome model expression, e.g., as autoregressive or moving average terms. The exogenous effects, similar to previous sections, can be seasonal trends and shocks either on product level or on product-supplier level. All the suggested main effects for the outcome model are illustrated on Table~\ref{tab:decspecs}. Finally, the additive noise terms are in all cases normally distributed random variables.

\subsubsection*{Simulation}

Operational decisions and the relative outcomes on a given day $\ell$ can be simulated using Algorithm~\ref{algo:dynamics}. The elements $\mathcal{I}(\ell),d_{\ell}^{-},\widetilde{y}_{\ell}$, $\widetilde{X}_{\ell}^{(d)}$, $\widetilde{X}_{\ell}^{(y)}$, $\mathcal{D}_{\ell}$, $\theta_{\ell}^{(0)}$, $\theta_{\ell}^{(1)}$, $\delta_{\ell}$, $\Sigma$, $f_{d,\ell}^{(0)}$, $f_{d,\ell}^{(1)}$, described in the previous sections are needed as input to generate an instance $(d_{\ell},y_{\ell})$ of decisions and outcomes for the day $\ell$.

\begin{table}
\small
{
\begin{algorithm}[H]
    \label{algo:dynamics}
    \KwIn{$\mathcal{I}(\ell)$, $d_{\ell}^{-}$, $\widetilde{y}_{\ell}$, $\widetilde{X}_{\ell}^{(d)}$, $\widetilde{X}_{\ell}^{(y)}$, $\mathcal{D}_{\ell}$, $\theta_{\ell}^{(0)}$, $\theta_{\ell}^{(1)}$, $\delta_{\ell}$, $\Sigma$, $f_{d,\ell}^{(0)}$, $f_{d,\ell}^{(1)}$}
    \KwOut{$(d_{\ell},y_{\ell})$}
    \textbf{Initialisation:} $d_{\ell}\leftarrow y_{\ell}\leftarrow\emptyset$\;
    \ForEach{$(l,m)$ \textbf{\emph{in}} $\mathcal{I}(\ell)$}{
        $U_{lm}^{(d,1)}(d)\leftarrow\left(f_{d,lm}^{(1)}(d,d_{\ell},d_{\ell}^{-})^{\top},\widetilde{X}_{\ell}^{(d,1)}(d)^{\top}\right)\theta_{\ell}^{(1)}$\;
        Draw $d_{lm}^{(0)}$ from $\text{Bernoulli}(U_{lm}^{(d,0)})$\;
        \If{$d_{lm}^{(0)}=0$}{Push $(0,\emptyset)$ to $d_{\ell}$\;\textbf{continue\;}}
        Remove index $(l,m)$ from $\mathcal{I}(\ell)$\;
        \ForEach{$d\in\mathcal{D}_{\ell}$}{
            $U_{lm}^{(d,1)}(d)\leftarrow\left(f_{d,lm}^{(1)}(d,d_{\ell},d_{\ell}^{-})^{\top},\widetilde{X}_{\ell}^{(d,1)}(d)^{\top}\right)\theta_{\ell}^{(1)}$\;
            $p^{(1)}(d)\leftarrow\exp\left(U_{lm}^{(d,1)}(d)\right)\Big/\sum_{d'\in\mathcal{D}_{\ell}}\exp\left(U_{lm}^{(d,1)}(d')\right)$\;
        }
        Draw $d_{lm}^{(1)}$ from Mult$(p^{(1)})$ and $\epsilon_{lm}^{(y)}$ from $\mathcal{N}\left(0,\Sigma\right)$\;
        Push $(1,d_{lm}^{(1)})$ to $d_{\ell}$ and $\left(\widetilde{y}_{lm}^{\top},\widetilde{X}_{\ell}^{(y)\top}\right)\delta_{\ell}+\epsilon_{lm}^{(y)}$ to $y_{\ell}$\;
    }
\caption{Procurement operations mechanism.}
\end{algorithm}}
\end{table}

\section{Numerical experiment}\label{sec:cstud}

One strength of the proposed framework is that it can simulate many possible scenarios by providing the configuration. A component of the system does not have knowledge of the working models of other components but only has access to historical records which are used to construct the features in its formal expression(s). We illustrate a use case of the proposed framework to compare supplier selection policies in the spot market. Our open sourced GitHub repository\footnote{\url{https://github.com/georgios-vassos1/procurement-ops}} can be used to reproduce all the results presented in this section.

Consider the problem of comparing the performance of different supplier selection policies over a year of operations. The user must specify the time-horizon of interest, the number of operational sites, and expressions of the time-to-request, requisition generation, decision-to-order, and outcome models. Moreover, a list of supplier utility models needs to be provided to induce the corresponding supplier selection policies. For instance, suppose that stakeholders in a maritime procurement business speculate over the amount of savings that can result from different supplier selection policies. The stakeholders may wish to support their claims with simulation-based evidence.

In this example, we consider a maritime procurement business in a container logistics company that operates 50 company-owned container ships for ocean transportation. The ships generate demand over a period of 365 days and may choose between 3 products when making a requisition. Purchase management can order the products from the spot market consisting of 2 suppliers. The stakeholders are concerned with fluctuations in the spot market and want to investigate whether the benefit of adopting dynamic supplier selection policies instead of issuing a static policy to be executed over the year. In the following, we describe the system configuration in detail and compare the performance of five different supplier selection policies in terms of regret, that is, a measure of divergence from the optimal policy.

\subsection{Demand model}

\subsubsection{Time-to-request model specifications}

The purpose of the intensity process model is to capture the heterogeneity patterns in requisition generation times. Heterogeneity can be due to temporal phenomena, e.g., seasonal trends, market shocks, etc, or due to differences between individual ships, e.g., number of crew members, model of the ship, size of the ship, scheduled port calls, etc. In this simple numerical study, we assume that requisition generation times are homogeneous both in time and between individuals. In particular, we set $\lambda_{i}(t)=\lambda=90$ days for all ship indices $i=1,\dots,50$. Requisition generation times are, thus, sampled from an exponential distribution with mean $1/\lambda$.

In the proposed discrete-event simulation design, the first requisition generation time is sampled for each ship during initialisation. Every time the requisition generation routine is invoked to create a requisition for a given ship, it will sample the next requisition generation time for that ship.

\subsubsection{Requisition generation model specifications}

We have seen in previous sections that upon triggering a request a requisition must be sampled from the generation model described in Sections \ref{sec:212} and \ref{sec:32}. In Section \ref{sec:212}, we also mentioned a first-stage propensity-to-stop model which effectively controls the size of the requisitions. For $\widetilde{e}_{ijk}^{(01)}$ and $\widetilde{e}_{ijk}^{(02)}$, refer to Table~\ref{tab:d2specs}, we consider the nonlinear effect of the requisition size $-0.1(k-1)^{2}$ and the baseline cost estimate of the requisition before adding a $k$-th line item. The latter is set to $0$ if the current cost estimate is less than \$20 and, at a sublinear rate, falls to $-120$ at \$500. The baseline cost estimate for any of the 3 products is fixed to \$100, \$50, and \$50. A linear expression of this two features and a random intercept is used in the logistic distribution to compute the propensity-to-stop score. The random intercept is sampled once at the start of the simulation from a normal distribution with mean 5 and variance 1 and is fixed thereafter. The decision-to-stop is a draw from a Bernoulli distribution with success rate equal to the propensity-to-stop score. The product utility and quantity intensity models are a normal distribution with mean equal to 0 and variance 1 and a Poisson distribution with means 0.1, 0.5, and 0.5, respectively.

It is emphasised that the proposed simulation framework can readily accommodate heterogeneity patterns in the product utility and quantity intensity models. However, to keep this numerical example simple, we disregard this layer to avoid unnecessary layers of complexity. The focus of this numerical study is to compare supplier selection policies in a two-supplier dynamic sport market. An alternative scenario could be to investigate the effect of heterogeneous demand on the total savings of the business under a particular supplier selection policy. The proposed simulation framework can be used to address a plethora of business inquiries.

\subsection{Decision and outcome models}

\subsubsection{Decision-to-order model specifications}

The propensity-to-order model described in Sections \ref{sec:22} and \ref{sec:33} is also kept simple to avoid complicating the interpretations of results in this numerical example. We sample the propensity-to-order score every time, for each item, in the list of unresolved requests, from a uniform distribution in $[0.1,0.9]$. This decision could be particularly interesting in a combined contract-spot market environment where additional costs might be incurred if the quantity commitment is not allocated to the contracted suppliers within the relative time frame.

\subsubsection{Outcome generation model specifications}

We consider a 3-dimensional outcome variable containing the unit price, delivery time and quality cost-equivalent value supposedly in USD; clearly, the cost-equivalent of delivery time is an increasing function, whereas that of quality is a decreasing function. The latter is due to the fact that at a given price higher quality entails a lower quality-cost. In this example, the outcome model has a VAR$(2)$ (vector autoregressive of order 2) structure, a short-term (past 90 days) allocation effect, a long-term (past 365 days) allocation effect, and an annual harmonic with initial phase set to $\pi/6$. The allocation effects correspond to the total quantity ordered from the selected supplier over the relative period of time.

\subsubsection{Supplier utility model specifications}

At this point, having specified all the above components, we can simulate the evolution of the procurement system under different supplier selection policies. In particular, we compare five different policies, namely, always selecting suppler index 1 (Supplier.1), always selecting supplier index 2 (Supplier.2), selecting at random (Random), selecting based on a random utility model (Utility), and selecting using a contextual mutli-armed bandit model (Bandit). The random utility model resembles a static supplier selection policy in that it is parameterised by an Euclidean vector that is fixed throughout the simulation. It is worth noting that the random utility and bandit models have strongly misspecified knowledge of the outcome generation mechanism based on solely observing the seasonal effect out of the many that influence the outcome dynamics.

The success measure that is used to compare the different policies is the regret and is defined as the difference between the optimal outcome and the outcome observed under the effective policy. The optimal outcome is computed by a working oracle that has knowledge of the true outcome distribution and, thus, always makes the optimal selection. The parameters of the bandit model are updated using the well-known Bayesian regression rule \citep[Part~IV]{gelman2004a}. The exploration-exploitation behavior of the bandit is regulated using Thompson Sampling \citep{Russo2017}. This is very important for the bandit to keep adapting to the changing dynamics of the spot market environment.

The bandit maintains a utility model parameterised by an Euclidean vector $\theta_{\ell}^{\mathit{(cmab)}}\in\mathbb{R}^{36}$ which stacks the parameters of all product-supplier outcome models. In more detail, for each of the 3 products there are two arms (supplier alternatives) each with an expected cost, lead time, and quality outcome. Each of the 18 stacked models is affiliated 2 parameters for the intercept and the observed seasonal effect. The information that must be available to compute the supplier utility with the bandit model is termed the context. Generally, the context can be understood as the summary measures
\begin{equation*}
    \widetilde{d}_{lm}^{(1)}=\left(f_{d,lm}^{(1)}(d,d_{\ell},d_{\ell}^{-})^{\top},\widetilde{X}_{\ell}^{(d,1)}(d)^{\top}\right)^{\top}
\end{equation*}
The distribution of $\theta_{\ell}^{\mathit{(cmab)}}$ is initialised to the multivariate normal distribution with mean $0^{(36\times1)}$ and variance equal to $70\cdot I^{(36\times36)}$, where $0^{(36\times1)}$ is a vector of 36 zeros and $I^{(36\times36)}$ is the $36\times36$ identity matrix. Below we summarise the logic of Thompson Sampling for a given line item $(l,m)$. It requires as input the context variable $\widetilde{d}_{lm}^{(1)}$ and current distribution of $\theta_{\ell}^{\mathit{(cmab)}}$, denoted $P_{\theta}$, and can readily replace lines 11 and 12 in Algorithm~\ref{algo:dynamics}. The steps are:
\begin{enumerate}
    \item Draw $\theta_{\ell}^{\mathit{(cmab)}}$ from $P_{\theta}$
    \item For each $d\in\{1,2\}$ set $U_{lm}^{(d,1)}(d)$ to $\widetilde{d}_{lm}^{(1)\top}\theta_{\ell}^{\mathit{(cmab)}}$
    \item Set $d_{lm}$ to the value $d$ that maximises $U_{lm}^{(d,1)}(d)$
    \item Update $P_{\theta}\,|\,u_{0}(y_{lm},d_{lm})$
\end{enumerate}
The regret corresponding to $(l,m)$ is easily computed by subtracting $U_{lm}^{(d,1)}(d)$ from $u_{0}(y_{lm},d_{lm})$. The utility is set to the inner product of the true outcome and a weighting rule; in this example, the weights are fixed to $(0.5,0.25,0.25)$. This combination is needed to convert the 3-dimensional outcome to a single welfare measure. The weight vector can be seen as a hyperparameter and can either be tuned with data or chosen based on domain expert knowledge.

\begin{figure}
  \centering
  \subfigure[]{%
  \resizebox*{7cm}{!}{\includegraphics{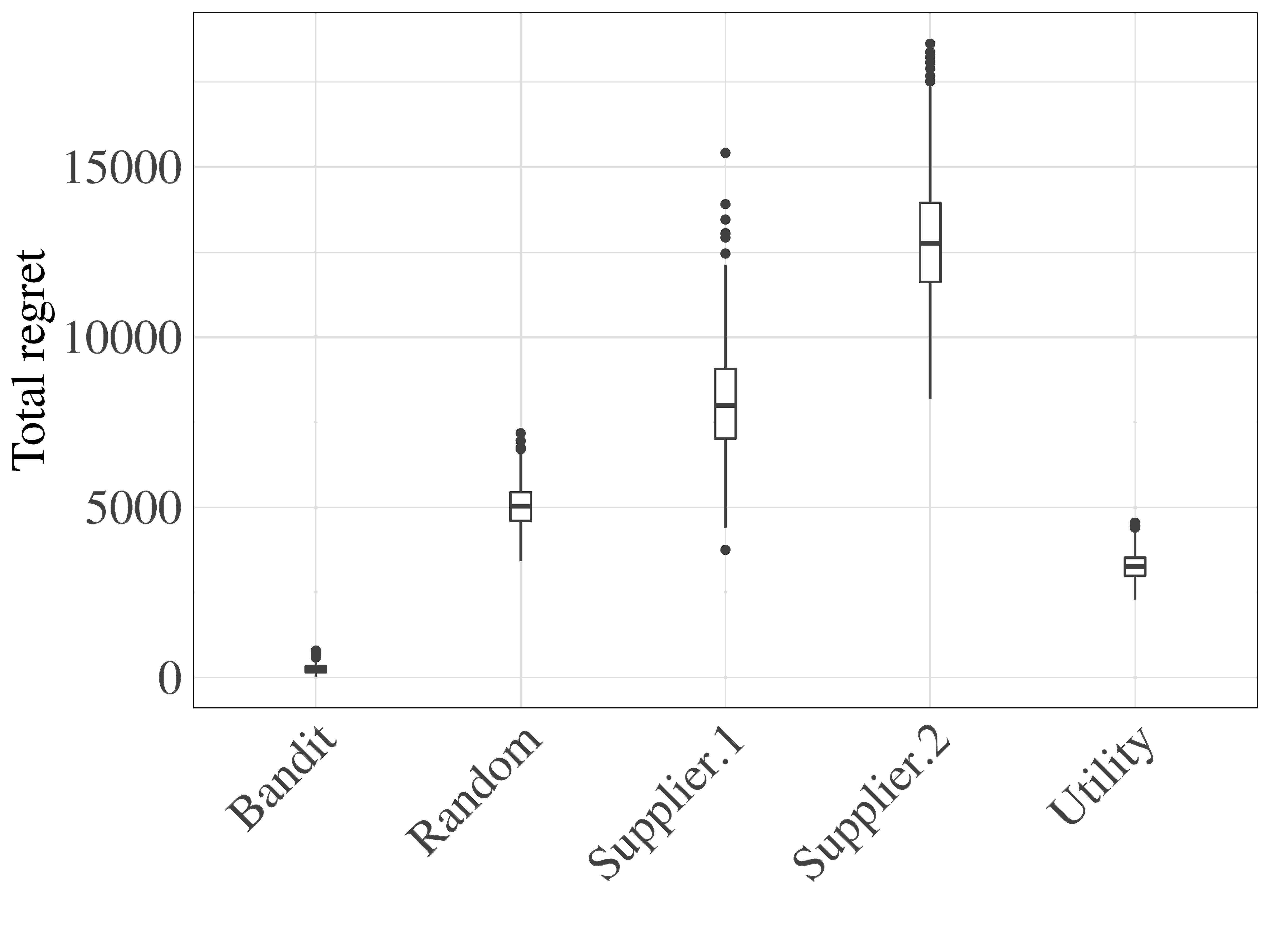}}}\hspace{5pt}
  \subfigure[]{%
  \resizebox*{7cm}{!}{\includegraphics{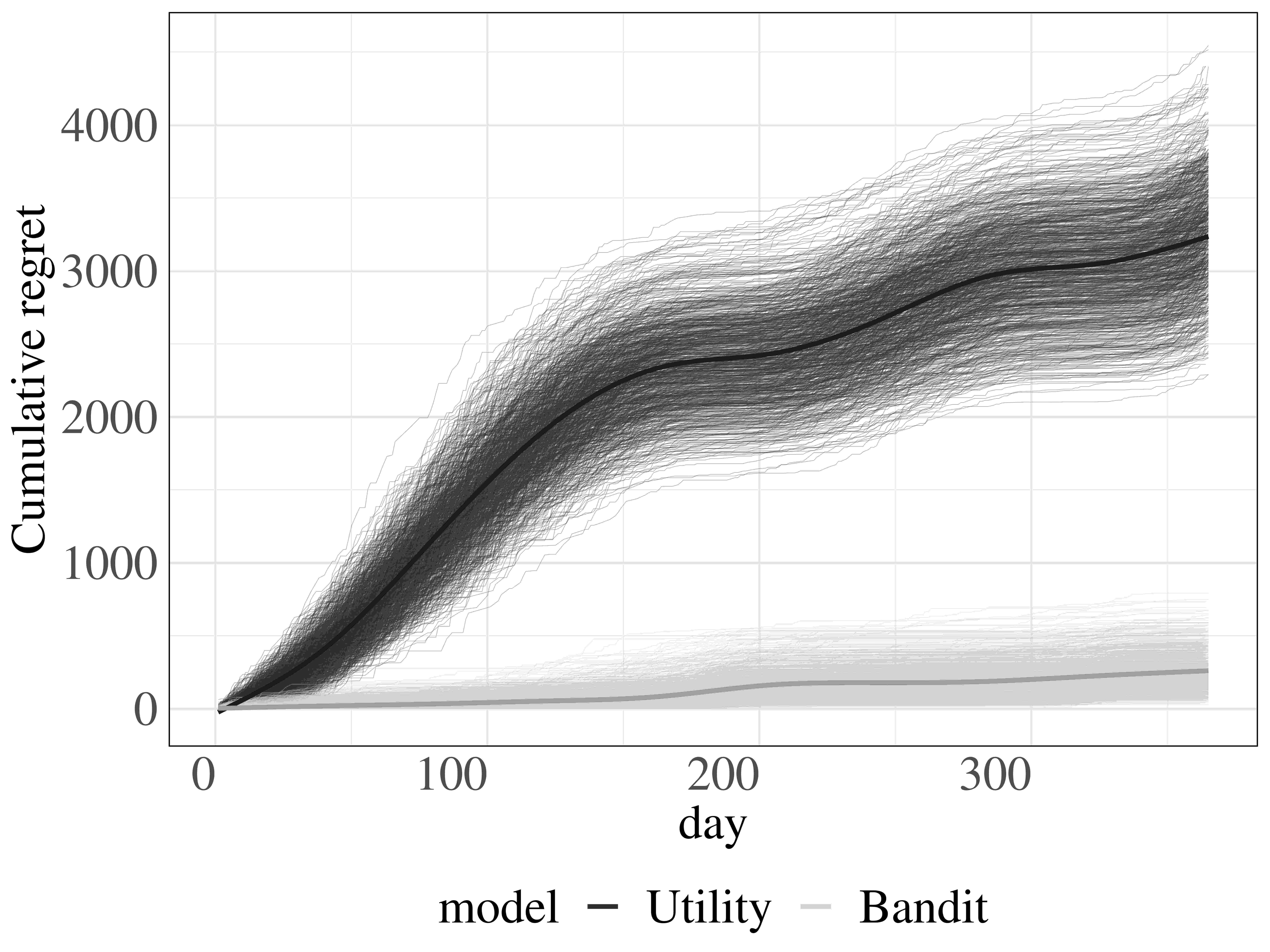}}}
  \caption{(a) Box plots of the total regret accumulated under five different supplier selection policies. Random policy means choosing a supplier at random, Supplier.1 and Supplier.2 correspond to always choosing the supplier with id 1 or 2, respectively, Utility is the random utility maximiser and Bandit is the contextual multi-armed bandit with access to the same information as the utility maximiser. (b) Spaghetti plots of the cumulative regret trajectory of the random utility maximising and contextual bandit policies.}
  \label{fig:results}
\end{figure}

We perform 1000 Monte Carlo simulations of the system dynamics, under the configuration described in this section, for every one of the five competing supplier selection policies. As expected, the random utility and bandit models had better performance than the three naive benchmarks. By and large, contextual multi-armed bandits have outperformed the rational utility maximisers, thus, confirming that data-driven decision-makers are beneficial in dynamic environments compared to their static alternatives. Figure~\ref{fig:results} (a) illustrates the terminal regret values incurred under the five different policies, whereas Figure~\ref{fig:results} (b) depicts the cumulative regret values incurred daily during the study period. Clearly, contextual multi-armed bandits are quite robust both to severe lack of knowledge of the true outcome distribution but also to the dynamics of the environment. We believe that many such advantages are to be expected from shifting to advanced machine learning methods.

\section{Conclusion and future works}\label{sec:concl}

This study develops a flexible framework for the simulation of procurement operations with a probabilistic model of operations and the generation of demand. Based on the lack of such unified modelling approach in procurement and given the success of PODS in airline revenue management, we tailor the logic of PODS to the procurement business. We have distinguished three main components of the evolutionary mechanism, namely, demand, decision, and outcome models, and described how data are generated by providing modelling choices and their assumptions. The nature of demand is given as a two-stage mechanism including a time-to-request and a requisition generation component. Models are proposed for the two most important operational decisions and three outcomes of interest in procurement. A discrete-event simulation logic is suggested to inform a software implementation of our simulation framework. Specific suggestions are made about input effects, modelling expressions, and algorithms to enable the implementation of our models using software. Finally, the simulation is used in a numerical experiment to investigate the performance of different supplier selection policies. Our results show that data-driven decision-makers, in particular, contextual multi-armed bandits, outperform rational utility maximisers in a dynamic spot market.

We believe that our effort can serve as the starting point for more statistical learning oriented research in the optimisation of decision-making in procurement systems, similar to PODS has led in the field of airline revenue management. In particular, many study cases can be made for tuning the parameters of our proposed models with observational data, e.g., learning the parameters of the demand model to understand the demand generation mechanism, the nature of demand can be of interest, by itself, for optimising inventory policies in warehouses used by the procurement business, and, as argued repeatedly in this study, to optimise the two key operational decisions, namely, when to place an order and supplier selection. However, to optimise the operational decisions a working model of the outcome is needed, thus, another interesting study case could be based on learning the parameters of the outcome model from observational data with continuous or categorical outcomes and potentially with some additional correlation structure. Learning the nature of the outcome mechanism is essential to the optimisation of when-to-order and supplier selection policies, we also tried to touch upon this topic through our insight in the case study. Furthermore, optimal policy learning and decision optimisation studies can be made by using postulated or estimated configurations of the demand and outcome models. Notice that, as with PODS, the decision-making component does not have access to the outcome model that is actually at work in the system but has access to a cost or reward model which induces a decision policy. For instance, one can start with an initial configuration of the decision-maker's cost model and apply online learning to obtain an optimal policy. Moreover, the estimation of an initial policy from observational data is also a case that can definitely be of interest. We conclude by stressing once more that with this work it is our hope to spark future research in procurement towards more flexible solutions, similar to PODS in the airline revenue management.

\section*{Notation}

In this section we explain some notation conventions that we have used to obtain more compact expressions. Due to the fact that we are concerned with an evolutionary system our observations have a panel structure of the sort
\begin{equation*}
    Z_{1},Z_{2},\dots,Z_{\tau}
\end{equation*}
when factorizing the probability density expression of such an observation we use the following convention
\begin{align*}
    p(Z_{1},\dots,Z_{\tau})&=\prod_{i=1}^{\tau}p(Z_{i}\,|\,Z_{1},\dots,Z_{i-1})\\
    &=\prod_{i=1}^{\tau}p(Z_{i}\,|\,\overline{Z}_{i-1})
\end{align*}
with $\overline{Z}_{0}=\emptyset$. When there are more than two components of interest in $Z$, e.g., $Z\doteq(X,A,Y)$, in temporal order, then we write
\begin{align*}
    \prod_{i=1}^{\tau}p(Z_{i}\,|\,\overline{Z}_{i-1})&=\prod_{i=1}^{\tau}p(X_{i},A_{i},Y_{i}\,|\,X_{i-1},A_{i-1},Y_{i-1})\\
    &=\prod_{i=1}^{\tau}p(Y_{i}\,|\,\overline{X}_{i},\overline{A}_{i},\overline{Y}_{i-1})\,p(A_{i}\,|\,\overline{X}_{i},\overline{A}_{i-1},\overline{Y}_{i-1})\,p(X_{i}\,|\,\overline{X}_{i-1},\overline{A}_{i-1},\overline{Y}_{i-1})\\
    &=\prod_{i=1}^{\tau}p(Y_{i}\,|\,Y_{i}^{-})\,p(A_{i}\,|\,A_{i}^{-})\,p(X_{i}\,|\,X_{i}^{-})
\end{align*}
where $Y_{i}^{-}\doteq(\overline{X}_{i},\overline{A}_{i},\overline{Y}_{i-1})$, $A_{i}^{-}\doteq(\overline{X}_{i},\overline{A}_{i-1},\overline{Y}_{i-1})$, and $X_{i}^{-}\doteq(\overline{X}_{i},\overline{A}_{i-1},\overline{Y}_{i-1})$. Furthermore, when we report a modelling expression we might want to construct a design matrix that could be enriched with interaction terms, basis expansions, and transformations of the main effects included, for instance, in $Y_{i}^{-}$. In such cases, we use a transformation mapping $f_{y,i}$ indexed both by the relative variable and by time, the latter because we could transform an expanding window of the past, and denote by $\widetilde{Y}_{it}$ the variable obtained from the mapping $Y_{i}^{-}\mapsto f_{y,i}(Y_{i}^{-})$.

Uppercase Latin letters are used to denote matrices, random variables, and simulated realizations in the pseudocode. Uppercase calligraphy Latin letters are used to denote sets. Lowercase Greek letters are used for unobserved effects and their simulated realizations in the pseudocode but also for configurable parameters. For features which are constructed from functions of observed and random variables, we use lowercase Latin characters with a tilde on top. In general, we use lowercase Latin characters for observed realizations of a random variable when the focus is on the generation mechanism.

\bibliographystyle{apalike}
\bibliography{refs.bib}

\end{document}